# AI-Driven Channel State Information (CSI) Extrapolation for 6G: Current Situations, Challenges and Future Research

Yuan Gao, Zichen Lu, Xinyi Wu, Wenjun Yu, Shengli Liu, Jianbo Du, Yanliang Jin, Shunqing Zhang, Xiaoli Chu, Shugong Xu

*Abstract*—CSI extrapolation is an effective method for acquiring channel state information (CSI), essential for optimizing performance of sixth-generation (6G) communication systems. Traditional channel estimation methods face scalability challenges due to the surging overhead in emerging high-mobility, extremely large-scale multiple-input multiple-output (EL-MIMO), and multi-band systems. CSI extrapolation techniques mitigate these challenges by using partial CSI to infer complete CSI, significantly reducing overhead. Despite growing interest, a comprehensive review of state-of-the-art (SOTA) CSI extrapolation techniques is lacking. This paper addresses this gap by comprehensively reviewing the current status, challenges, and future directions of CSI extrapolation for the first time. Firstly, we analyze the performance metrics specific to CSI extrapolation in 6G, including extrapolation accuracy, adaption to dynamic scenarios and algorithm costs. We then review both model-driven and artificial intelligence (AI)-driven approaches for time, frequency, antenna, and multi-domain CSI extrapolation. Key insights and takeaways from these methods are summarized. Given the promise of AI-driven methods in meeting performance requirements, we also examine the open-source channel datasets and simulators that could be used to train high-performance AI-driven CSI extrapolation models. Finally, we discuss the critical challenges of the existing research and propose perspective research opportunities.

## I. INTRODUCTION

The sixth generation (6G) of mobile networks are poised to deliver a transformative leap in connectivity and performance, and are anticipated to be deployed around 2030 [1]–[5]. 6G is expected to address a wide range of high-mobility communication scenarios [6] by using extremely large-scale MIMO (EL-MIMO) [7], [8] and higher frequency bands beyond the millimeter-wave (mmWave) spectrum [9]–[12]. Acquiring accurate channel state information (CSI) has been a key factor to ensure the performance of mobile networks, and is increasingly significant to achieve the vision of 6G [13]. However, a major challenge for 6G is to obtain accurate CSI while keeping the channel estimation overhead low [14], [15]. Conventional channel estimation relies on transmitting pilot symbols to measure channel responses, followed by techniques like interpolation or compressive sensing to reconstruct CSI [16], [17]. While effective in 5G massive MIMO, these methods falter in 6G's EL-MIMO and higher frequency contexts due to poor scalability, resulting in prohibitive overhead [18], [19]. This has spawned the birth of the research area of CSI extrapolation, which is an important technique in modern wireless communication systems, especially in the upcoming 6G era [20].

### A. An overview of CSI extrapolation

CSI extrapolation techniques aim to infer the complete CSI using a subset of CSI (mainly acquired using pilot-based channel estimation), thereby reducing the overhead [21]–[24]. Depending on the application, CSI extrapolation is categorized into several types, including time-domain, frequency-domain, and antenna-domain.

- In high-speed 6G environments, such as vehicle-to-everything (V2X) and drone networks, channels change rapidly due to high mobility. Traditional methods often struggle to maintain accurate CSI in these dynamic contexts, necessitating frequent pilot transmissions or feedback updates, which increases overhead significantly [22], [25]. By leveraging the temporal characteristics of channel states, time-domain CSI extrapolation addresses this by predicting future CSI from historical data, reducing frequent channel estimation and overhead. Current research primarily targets Time-Division Duplex (TDD) systems [26].

- Frequency-domain CSI extrapolation is vital for Frequency-Division Duplex (FDD) and multi-band systems [27], [28]. Partial reciprocity is observed in FDD systems, in which the delay and angle of UL and DL are almost equal, while the complex gains between DL and UL are distinct [29]. Frequency-domain CSI extrapolation can be utilized to exploit the partial reciprocity and extrapolate the UL channel using DL

This work was supported in part by Shanghai Natural Science Foundation under Grant 25ZR1402148, and in part supported by the 6G Science and Technology Innovation and Future Industry Cultivation Special Project of Shanghai Municipal Science and Technology Commission under Grant 24DP1501001. (Corresponding author: Yanliang Jin, Shugong Xu)

Yuan Gao, Zichen Lu, Xinyi Wu, Wenjun Yu, Shengli Liu, Yanliang Jin and Shunqing Zhang are with the School of Communication and Information Engineering, Shanghai University, China, email: gaoyuansie@shu.edu.cn, luzichen@shu.edu.cn, wu_xinyi0312@shu.edu.cn, yuwenjun@shu.edu.cn, victoryliu@shu.edu.cn, jinyanliang@staff.shu.edu.cn and shunqing@shu.edu.cn.

Jianbo Du is with the School of Communication and Information Engineering, X'an University of Posts and Telecommunications, X'an, China, e-mail: dujianboo@163.com.

Xiaoli Chu is with the Department of Electronic and Electrical Engineering, the University of Sheffield, UK, e-mail: x.chu@sheffield.ac.uk.

Shugong Xu is with Xi'an Jiaotong-Liverpool University, Suzhou, China, email: shugong.xu@xjtlu.edu.cn.

TABLE I
List of acronyms in alphabetical order

| Acronym | Explanation |
|---|---|
| 1G | 1th Generartion |
| 2D | Two-dimensional |
| 3D | Three-dimensional |
| 3GPP | 3rd Generation Partnership Project |
| 5G | 5th Generation |
| 6G | 6th Generation |
| ADEN | Antenna Domain Extrapolation Network |
| AGMAE | Asymmetric Graph Masked Autoencoder |
| AI | Artificial Intelligence |
| AoA | Angle of Arrival |
| AoD | Angle of Departure |
| AR | Autoregressive |
| ASN | Antenna Selection Network |
| BEE | Basis Expansion Extrapolation |
| BEM | Basis Expansion Modeling |
| BER | Bit Error Rate |
| BS | Base Station |
| CCM | Channel Correlation Metric |
| CDL | Clustered Delay Line |
| CENet | CSI extrapolation Network |
| CFR | Channel Frequency Response |
| CIR | Channel Impulse Response |
| CNN | Convolutional Neural Network |
| CPcGAN | Conditional Generative Adversarial Network |
| CSI | Channel State Information |
| CT | Channel Transformer |
| CVNN | Complex-valued Neural Network |
| D2D | Device-to-Device |
| DFT | Discrete Fourier Transform |
| DL | Downlink |
| DNN | Deep Neural Network |
| EM | Extrapolation metric |
| ESPRIT | Estimation of Signal Parameters via Rotational Invariance Techniques |
| FDD | Frequency Division Duplex |
| FIT | First-order Taylor Expansion |
| FNN | Fully Neural Network |
| GA | Ground-to-Air |
| GAI | Generative Artificial Intelligence |
| GAN | Generative Adversarial Network |
| GRU | Gated Recurrent Unit |
| HRPE | High-resolution Parameter Estimation |
| I2O | Indoor-to-Outdoor |
| IENet | Interference Elimination Network |
| IIoT | Industrial Internet of Thing |
| IoT | Internet of Thing |
| ITU | International Telecommunication Union |
| JCPAS | Joint Channel Prediction and Antenna Selection |
| KF | Kalman Filter |
| LB | Lower Bound |
| LEO | Low Earth Orbit |
| LMMSE | Linear Minimum Mean Squared Error |
| LoS | Line-of-Sight |
| LS | Least Squares |
| LSTM | Long Short-Term Memory |
| LTE | Long Term Evolution |
| MCGP | Monte Carlo Gaussian Process |
| MDMP | Multidimensional Matrix Pencil |
| mDRUNet | Modified Deep Residual U-Shaped Network |
| MICP | Mobility Induced Channel Prediction |
| MIMO | Multiple-input Multiple-output |
| MLP | Multilayer Perceptron |
| MMSE | Minimum Mean Squared Error |
| mmWave | Millimeter Wave |
| MPC | Multipath Component |
| MSE | Mean Square Error |
| MUSIC | Multiple Signal Classification |
| NLoS | Non-Line-of-Sight |
| NMSE | Normalized Mean Square Error |

| Acronym | Explanation |
|---|---|
| NR | New Radio |
| O2I | Outdoor-to-Indoor |
| O2O | Outdoor-to-Outdoor |
| ODE | Ordinary Differential Equation |
| QoS | Quality of Service |
| RIS | Reconfigurable Intelligent Surface |
| RMa | Rural Macro |
| RMSE | Root Mean Squared Error |
| RNN | Recurrent Neural Network |
| RSSI | Received Signal Strength Indicator |
| SCNet | Sparse Complex-Valued Neural Network |
| SCP | Spatial Consistency Property |
| SE | Spectral Efficiency |
| SISO | Single-input Single-output |
| SNR | Signal-to-Noise Ratio |
| SVR | Support Vector Regression |
| TAS | Transmit Antenna Selection |
| TDD | Time Division Duplex |
| TDL | Tapped Delay Line |
| TDoA | Time Difference of Arrival |
| THz | Terahert |
| TL | Transfer Learning |
| TR | Technical Report |
| UAV | Unmanned Aerial Vehicle |
| UCB | Upper Confidence Boundary |
| UE | User Equipment |
| uGP | Uncertain Gaussian Process |
| UL | Uplink |
| UMa | Urban Macrocell |
| UMi | Urban Microcell |
| UT | User Terminal |
| V2V | Vehicle-to-Vehicle |
| V2X | Vehicle-to-Everything |
| VSS | Vector Spatial Signature |
| WTMP | Wavefront Transform Matrix Pencil |

channel or vice versa [30]. Multi-band systems generally utilize multiple spectrum bands, such as the sub-6 GHz and mmWave or THz bands. The mmWave and THz bands exhibit unique propagation properties, such as higher path loss and greater directionality, resulting in sparser channels [31]. These sparse characteristics could be captured by the frequency-domain CSI extrapolation to extrapolate the CSI of unobserved bands using observed bands.

- The sheer scale of EL-MIMO systems increases the number of CSIs requiring estimation dramatically increase the need for CSI estimation per antenna. Conventional practice involves dedicating a pilot symbol for each antenna, which becomes inefficient as the number of antennas grows, consuming substantial time-frequency resources [32]. Antenna-domain CSI extrapolation proposes using part of the antennas' CSI to predict others, leveraging spatial correlations between channel states [33].

Beyond single-domain extrapolation, the increasing complexity of 6G scenarios, such as an ultra-high speed communication served by en EL-MIMO system, necessitates a multi-domain CSI extrapolation [34], [35]. Multi-domain joint CSI extrapolation involves inferring unknown CSI across multiple dimensions of time, frequency and antenna based on known CSI, exploiting the correlation of CSIs across different domains to improve the accuracy of channel estimation [36], [37].

TABLE II

Summary of existing surveys and magazines related to CSI extrapolation. ●, ○and ◐indicate the topic is well-covered, partial-covered and not covered, respectively.

| Paper | Year | Time-domain | Freq-domain | Ant-domain | Multi-domain | Datasets | Key insights & limitations |
|---|---|---|---|---|---|---|---|
| [38] | 2024 | ● | ○ | ○ | ○ | ○ | Mainly reviewing RNN-series-based CSI extrapolation approaches in time-domain, neglecting model-driven and advanced AI-based approaches. |
| [39] | 2025 | ● | ○ | ○ | ○ | ○ | Reviewing both model-based and AI-based CSI extrapolation approaches in time-domain, without in-depth discussion of generative AI, neglecting other domains. |
| [21] | 2021 | ○ | ● | ● | ○ | ○ | Elaborating on the principles of frequency and antenna-domain CSI extrapolation, focusing on variations of CNN and DNN, without reviewing advanced AI models. |
| [30] | 2023 | ● | ● | ● | ○ | ○ | Discussing the principles of time, frequency and antenna-domain CSI extrapolation,focusing on AI-based models, without reviewing multi-domain CSI extrapolation. |
| [40] | 2023 | ● | ○ | ○ | ○ | ◐ | Comparing AI-based time-domain CSI extrapolation approaches using the channel data measured in the proposed testbed, neglecting other domains. |
| [41] | 2023 | ◐ | ○ | ○ | ○ | ○ | Mainly reviewing model-driven and RNN-series-based CSI extrapolation approaches in time-domain, neglecting advanced AI-based approaches and other domains. |
| [42] | 2019 | ● | ● | ○ | ○ | ○ | Reviewing basic AI-based (e.g., DNN and CNN) CSI extrapolation approaches in time and frequency-domain, advanced AI-based approaches and other domains not covered. |
| This paper | | ● | ● | ● | ● | ● | Comprehensively reviewing the model-based and AI-based CSI extrapolation approaches in time, frequency, antenna, and multi-domain, with in-depth discussion of their cons and pros. Reviewing the existing available datasets and channel simulators. Discussion of the challenges and future research. |

## B. Related Surveys on CSI extrapolation

There have been several surveys on CSI extrapolation focusing on a single domain CSI extrapolation [38], [39], [43]. [38] focuses on time-domain channel prediction using data-driven neural networks (NNs) under the standardized 3rd generation partnership project (3GPP) tapped delay line (TDL)-A model. It rigorously compares five NN architectures, including multilayer perceptron (MLP) [44], convolutional neural network (CNN) [45], long short-term memory (LSTM) [46], gated recurrent units (GRU) [47], and Transformers [48], in terms of prediction accuracy, robustness to channel aging, and computational complexity. [39] presents a broader framework for artificial intelligence (AI)-based channel prediction, categorizing approaches into time-domain channel prediction (e.g., auto-regressive (AR) models, Kalman filters, NNs) and environmental adaptation (e.g., transfer learning, meta-learning, data augmentation). The above surveys primarily discuss time-domain channel, without discussion on frequency-domain, antenna-domain, or joint multi-domain CSI extrapolation. [43] makes significant contributions to antenna-domain channel generation and extrapolation in mobile communications, particularly for massive MIMO systems. It introduces conditional diffusion models (DM) as a novel generative AI framework to address challenges in high-dimensional antenna-domain channel estimation, extrapolation, and feedback.

Several paper reviewed CSI extrapolation in more than one domains. [21] reviewed the deep learning (DL) solutions for antenna and frequency-domain CSI extrapolation, such as deep neural network (DNN) and CNN-based models. [30] systematically explored CSI extrapolation techniques across time, frequency, and antenna-domain to address overhead challenges in 6G systems. For time-domain extrapolation, the authors emphasized the spatial consistency property (SCP) of channels in high-mobility scenarios, proposing generative adversarial networks (GAN)-based networks to handle nonlinear CSI evolution. In frequency-domain extrapolation, they distinguished between FDD uplink (UL)-to-downlink (DL) mapping and multiband coexistence (e.g., sub-6 GHz to mmWave), leveraging partial reciprocity and diffraction phenomena for parameter-level inference. For antenna-domain extrapolation, a novel channel Transformer with self-attention mechanisms is introduced to address spatial non-stationarity in EL-MIMO, enhanced by transfer learning from time-domain pretraining. [41] reviewed AI-based channel quality prediction techniques, emphasizing their role in reducing pilot signal overhead for 6G networks. It categorized approaches into time, frequency and antenna-domain. Time-domain prediction employed recurrent neural network (RNN) to exploit temporal coherence, while frequency-domain methods used DNNs to map UL/DL correlations. Antenna-domain CSI extrapolation is achieved by leveraging k-nearest neighbors (KNN) and CNN for radio map completion, and network correlation-based prediction introduced a DNN framework to infer direct device-to-device (D2D) channels from reference node measurements. However, the work overlooked advanced techniques like Transformer-based models or hybrid domain fusion strategies. [42] provided an in-depth analysis of RNN-based channel prediction techniques for adaptive wireless communication systems. It began by critiquing traditional model-based approaches, such as parametric and AR models, highlighting their limitations in computational complexity (parametric models) and susceptibility to noise (AR models). The authors then introduced RNNs as a data-driven alternative, emphasizing their capability to exploit temporal correlations in fading channels. The paper outlined the architecture and training methodology of RNNs, including back-propagation through time, and extends

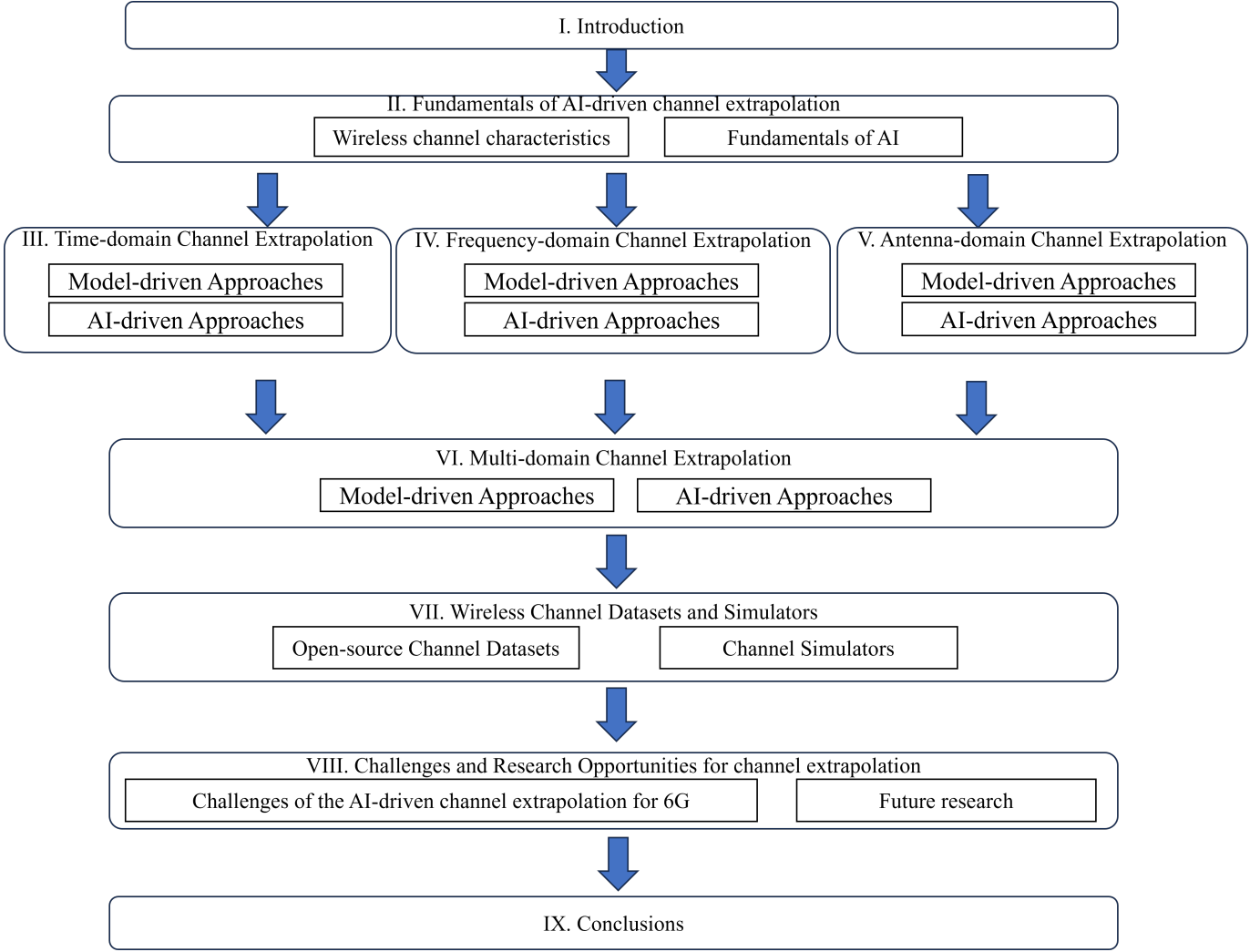

Fig. 1. Outline of this paper.

their application to flat-fading and frequency-selective MIMO-OFDM systems.

However, the above surveys have the following limitations:

- The above surveys provide high-level overviews of the techniques for CSI extrapolation, but lack a systematic comparison of different extrapolation techniques, assessing their strengths, weaknesses, and suitability for various scenarios.
- The above surveys focus on the CSI extrapolation in specific domain(s), such as time-domain or frequency-domain extrapolation, for example [38], [39] only consider time-domain extrapolation, while [43] only focuses in antenna-domain extrapolation. However, none of the above surveys review the research status of multi-domain CSI extrapolation.
- The above surveys mainly focus on the AI-based CSI extrapolation schemes, neglecting the discussion on the available datasets and simulators. It is well recognized that high-quality dataset play a key role to train high-performance AI-based models.

### C. Scope and Organization

This paper aims to comprehensively discuss the current status, challenges, and solutions in CSI extrapolation. To the best of our knowledge, this is the first survey offering an in-depth analysis of time, frequency, antenna, and multi-domain CSI extrapolation. The novelty and contributions of this work can be summarized as follows:

- **Foundation of AI-driven CSI extrapolation**: We present a comprehensive review of the foundational knowledge of both CSI extrapolation and AI technologies. For CSI extrapolation, we systemically formulate the problem of CSI extrapolation and analyzed the typical use cases in time, frequency, antenna and multi-domain in 5G and 6G. On the AI side, we review representative AI models in terms of principles, application for CSI extrapolation and computational complexity.
- **Comprehensive review of CSI extrapolation schemes**: Unlike existing surveys that offer high-level overviews, we perform a detailed comparison of CSI extrapolation techniques. This includes a thorough examination of both model-driven and AI-driven schemes, highlighting their respective strengths and weaknesses. We conduct a comprehensive review of multi-domain CSI extrapolation, emphasizing its unique challenges as compared to single-domain extrapolation. To our knowledge, this is the first survey to examine multi-domain extrapolation in such detail.

- **Available datasets and channel simulators**: To support effective training of AI-driven models, we present an extensive survey of accessible open-source wireless datasets and channel simulators pertinent to CSI extrapolation. We analyze the key features, strengths and weaknesses of the datasets/simulators for CSI extrapolation. This aspect has been largely overlooked by previous survey and review paper.

- **Challenges and Future Directions**: We discuss the major challenges in AI-driven CSI extrapolation in time, frequency, antenna, multi-domain. We also present corresponding future research directions, including effective and reliable dataset construction, comprehensive performance evaluation metrics, advanced model design, and integrating with emerging techniques.

This manuscript is organized as follows. Section II discussed the performance metrics of CSI extrapolation for 6G, including the extrapolation accuracy, generalization and computational complexity. In section III, IV, V and VI, we reviewed the state-of the art (SOTA) model-driven and AI-driven approaches for time, frequency, antenna and multi-domain CSI extrapolation, respectively. Key insights and takeaways for the current research on time, frequency, antenna and multi-domain CSI extrapolation are summarized. As the AI-driven approaches are promising to meet the performance requirements of CSI extrapolation, we reviewed the wireless channel datasets and simulators that could be used to train high-performance AI models for CSI extrapolation in Section VII. Finally, in Section VIII we discussed several critical research challenges and provided potential research directions to resolve each challenge.

## II. Fundamentals of AI-driven CSI extrapolation

This section elaborates applications and use cases of CSI extrapolation, along with the fundamentals of representative AI models.

### A. Wireless channel characteristics

To illustrate the wireless channel characteristics, we consider a MIMO system using orthogonal frequency-division multiplexing (OFDM). The wireless channel between the $i$-th antenna at the BS and $j$-th antenna at the UE in frequency $f$ at time instance $t$ is given as:

$$h_{i,j}(f,t) = \sum_{l=1}^{L} \alpha_{i,j,l}(f,t) e^{j\eta_{i,j,l}(f,t)} e^{-j2\pi f_l(t)\tau_l(t)}, \quad (1)$$

where $\alpha_{i,j,l}(f,t)$ and $\eta_{i,j,l}(f,t)$ are the amplitude and the phase of the $l$-th path between the $i$-th antenna at the BS and $j$-th antenna at the UE in frequency $f$ at time instance

$t$, respectively. $L$ denotes the total number of paths. $f_l(t)$ and $\tau_l(t)$ are the receiving frequency and transmission delay of the $l$-th path, respectively. The complete channel matrix $\mathbf{H}(f,t)$ in frequency $f$ at time instance $t$ can be expressed as:

$$\mathbf{H}(f,t) = \begin{bmatrix} h_{1,1}(f,t) & ... & h_{1,j}(f,t) & ... & h_{1,J}(f,t) \\ ... & ... & ... & ... & ... \\ h_{i,1}(f,t) & ... & h_{i,j}(f,t) & ... & h_{i,J}(f,t) \\ ... & ... & ... & ... & ... \\ h_{I,1}(f,t) & ... & h_{I,j}(f,t) & ... & h_{I,J}(f,t) \end{bmatrix}, \quad (2)$$

where $I$ and $J$ are the number of antennas at the BS and the UE, respectively.

*1) Time-domain:* In conventional communication systems prior to 5G, due to the limited mobility of UEs, the channel varies slowly in time-domain, and could be considered as unchanged during a period of time, i.e., the coherence time, which could be illustrated mathematically as:

$$h_{i,j}(f, t + \Delta t) \approx h_{i,j}(f,t), \forall i, j, f, \quad (3)$$

where $\Delta t$ is the time difference between the samples of channel. Eq. (3) holds if the time difference $\Delta t$ is smaller than the coherence time $t_C$, i.e., $\Delta t < t_C$. To be noted that the coherence time $t_C$ is negatively related to the relative-mobility between the BS and UE, i.e., if the UE moves slowly, the coherence time is long. To guarantee the performance of cellular systems, CSI should be acquired (mainly via pilot-based channel estimation) during every coherence time, which indicates that the longer the coherence time, the less frequent channel acquisition can be performed, resulting in smaller overhead. However, 6G is expected to support a wide range of high-speed communication scenarios, such as V2X and drone networks, the coherence time becomes much smaller. This means that the acquired CSI will be outdated in a short time period, which is known as the phenomenon of channel aging or channel staleness. Therefore, the conventional pilot-based channel estimation is required to be performed more frequently, resulting in excessive overhead.

The essence of the channel variations in time-domain arises from the following 2 aspects:

- **Movement-induced Doppler frequency shift**: the relative movement between the BS and UE will cause Doppler frequency shift, which is propositional to the radical speed between the BS and UE. This means that the frequency of each path $f_l(t)$ will change, resulting a variation of the wireless channel given in Eq. 1. The more significant the Doppler frequency shift, the larger variation of wireless channel in a certain time period. In this respect, the time-domain CSI extrapolation for 6G is more challenging than that for 5G, as 6G is expected to support the mobility over 1000 km/h, which is much faster than that in 5G (500 km/h).

- **Relative location change between BS and UE**: the relative movement between the BS and UE will change the amplitude $\alpha_{i,j,l}(f_C, t)$ and phase $\eta_{i,j,l}(f_C, t)$ of the $l$-th path, this will inevitably change the wireless channel given in Eq. 1. For an even larger location change, the total number of paths $L$ is more likely to change. The faster the UE moves, the larger variation of wireless channel tends to occur in a certain time period.

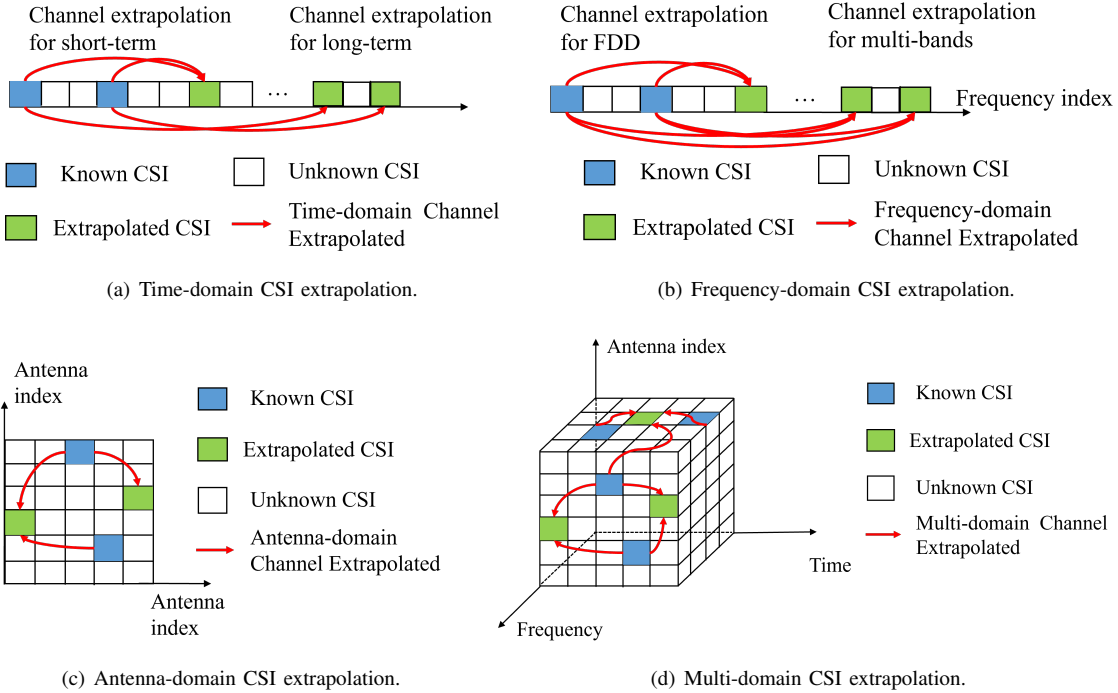

(a) Time-domain CSI extrapolation.

(b) Frequency-domain CSI extrapolation.

(c) Antenna-domain CSI extrapolation.

(d) Multi-domain CSI extrapolation.

Fig. 2. Illustration of CSI extrapolation.

TABLE III
Comparison of time-domain, frequency-domain, antenna-domain and multi-domain CSI extrapolation for 5G and 6G in terms of significance, applications & challenges.

| | 5G | 6G |
|---|---|---|
| Time-domain | Short-term: Doppler shift due to high speed; Long-term: Doppler shift and location change. | Short-term: Doppler shift due to ultra-high speed; Long-term: Doppler shift and location change. |
| Frequency-domain | FDD: partial channel reciprocity; Multi-band: channel charcteristics between sub-6GHz and mmWave in terms of dependence on LoS, atmospheric absorption, etc. | FDD: partial channel reciprocity; Multi-band: channel charcteristics between sub-6GHz (mmWave) and THz, in terms of dependence on LoS, atmospheric absorption, scattering, diffraction, etc. |
| Antenna-domain | Far-field: location change of antennas | Emerging MIMO systems: in addition to the location change of antennas, spatial nonstationarity in EL-MIMO, flexible location of antenna in fluid antenna systems, phase control in reconfigurable intelligent surface. |
| Multi-domain | Coupling of channel correlation among multiple domains. | Coupling of channel correlation among multiple domains, challenges of channel extrpolation in single domain are added up |

As illustrated in Fig. 2(a), time-domain CSI extrapolation is proposed to acquire the CSI in high-mobility scenarios using the historical CSI as:

$$[h_{i,j}(f,t), ..., h_{i,j}(f, t + t_{M-1})] \qquad (4)$$
$$= \mathbb{F}_{\mathrm{T}}\left([h_{i,j}(f, t - t_1), ..., h_{i,j}(f, t - t_N)]\right), \qquad (5)$$

where $[t, ..., t + t_{M-1}]$ and $[t - t_1, ..., t - t_N]$ are the $M$ time instances of the extrapolated CSI and the $N$ time instances of the historical, respectively. $\mathbb{F}_{\mathrm{T}}$ is the time-domain CSI extrapolation. This technique is particularly critical in mobile communications, especially in scenarios with rapidly changing channels (e.g., high-mobility users), where accurate CSI prediction can optimize resource allocation, beamforming, and precoding design. The challenging of short-period time domain extrapolation arises from the Doppler frequency shift, while for the long-period time domain extrapolation, both the Doppler frequency shift and location-change need to be considered.

*2) Frequency-domain:* FDD and multi-band systems are promising to achieve the stringent latency and capacity requirements of 6G. Specifically, compared with TDD, FDD systems are capable of supporting both the UL and DL communication simultaneously. The latency can be minimized by avoiding the transmission termination due to the UL-2-DL or DL-2-UL switch in TDD. Higher frequency bands, such as mmWave, THz bands, are proposed to enhance the capacity of cellular systems. Conventional pilot-based channel estimation for either FDD or multi-band systems will inevitably lead to enlarging overhead, reducing the network performance.

The essence of the channel variations in time-domain arises from the following 2 aspects:

- **FDD systems**: the UL and DL bands are separated in FDD, and the duplex spacing of bands between DL $f_{DL}$ and UL $f_{UL}$ is almost always much larger than the channel coherence bandwidth $f_C$, i.e., $|f_{DL} - f_{UL}| > f_C$. Thus, the channel reciprocity does not hold:

$$h_{i,j}(f_{UL}, t) \neq h_{i,j}(f_{DL}, t), \forall i, j, t. \qquad (6)$$

  Research observed that for some cases, the only difference between the DL and UL channels is the complex gain of each propagation path, while the delay and angle of the DL and UL channels are frequency independent [29]. Such partial channel reciprocity makes it possible to infer the DL channel using UL channel or vice versa.

- **Multi-band systems**: for a system operating in both the sub 6 GHz and mmWave (THz) band, the channel characteristics of these two bands are quite different. Specifically, the mmWave (THz) band is more sensitive to blockage than sub 6 GHz. In addition, mmWave and THz bands have unique propagation characteristics. mmWave supports reliable links up to hundreds of meters in line-of-sight (LOS) and limited non-line-of-sight (NLOS) via diffraction and reflection, with moderate oxygen and rain attenuation. In contrast, THz suffers extreme atmospheric absorption, negligible diffraction, severe scattering, and near-total dependence on LOS, confining practical wireless transmission to meters or less. The above propagation characteristics among different bands make frequency-CSI extrapolation extremely challenging.

As shown in Fig. 2(b), frequency-domain CSI extrapolation is proposed to acquire the CSI in one frequency band using the known CSI in another frequency band as:

$$[h_{i,j}(f'_1, t), ..., f'_M, t)] \qquad (7)$$
$$= \mathbb{F}_F \left( [h_{i,j}(f_1, t), ..., h_{i,j}(f_N, t)] \right), \qquad (8)$$

where $[f'_1, ..., f'_M]$ and $[f_1, ..., f_N]$ are the $M$ sampling frequency points of the target spectrum band(s) and the $N$ sampling frequency points of the spectrum band(s) with known CSI, respectively. $\mathbb{F}_F$ is the frequency-domain CSI extrapolation. This approach is particularly valuable in scenarios like FDD and multi-band systems, where reducing feedback overhead and enhancing spectral efficiency are critical. Partial channel reciprocity can be exploited for the frequency-domain CSI extrapolation in FDD systems. The challenge of the frequency-domain extrapolation arises from the distinct channel characteristics between multiple spectrum bands.

*3) Antenna-domain:* An unprecedented trend in the involution of cellular networks is that the number of antennas is increasing excessively, from 4 antennas in 4G to over 1024 in 6G. Conventional approach to acquire the CSI of MIMO systems is by allocating dedicated pilots to each Tx-Rx pair, which is affordable for 4G. However, such scheme is not practical for 6G, which will result in excessive overhead. The essence of channel variation in antenna-domain, i.e., for difference Tx-Rx pairs, arises from the following 2 aspects:

- **Far field**: in 5G the distance between BS and UE is much larger than the Rayleigh distance, the conventional far-filed assumption holds, which simplifies the wireless propagation by assuming spatial stationarity, i.e., the the number and angle of the multiple paths are identical. This indicates that variation of channel between different Tx-Rx pairs is caused by the location of the antennas.

- **Emerging antenna systems**: for EL-MIMO systems, the far-filed assumption is invalid, and the spatial nonstationarity becomes significant, which might lead to that some propagation paths are only visible to parts of the antennas of a massive antenna array [49], [50]. In this case, it turns that larger the separation between antenna locations, the greater the disagreement among channels in terms of the number of paths $L$, and thereby the amplitude $\alpha_{i,j,l}(f,t)$, phase $\eta_{i,j,l}(f,t)$ transmission delay $\tau_l(t)$. Other types of programmable antenna, such as fluid antenna, reconfigurable intelligent surface, etc, provide more flexible approaches to optimize the wireless channel via changing the location and phase of the antenna element, provide new challenges for antenna-domain CSI extrapolation.

As illustrated in Fig. 2(c), antenna-domain CSI extrapolation is proposed to acquire the CSI of all the antennas using the known CSI a subset of antennas as:

$$h_{\{i,j\}_I}(f,t) = \mathbb{F}_A(h_{\{i,j\}_K}(f,t)), \qquad (9)$$

where $\{i,j\}_I$ and $h_{\{i,j\}_K}(f,t)$ are the set of Tx-Rx pairs of interest and the set of Tx-Rx pairs with known CSI, respectively. To be noted that the number of Tx-Rx pairs of interest is generally much larger than the number of Tx-Rx pairs with known CSI for minimization of the overhead, i.e, $\|\{i,j\}_I\| \gg \|\{i,j\}_K\|$. $\mathbb{F}_A$ is the antenna-domain CSI extrapolation, which aims to infer the CSI of unmeasured antennas on the same panel using CSI from a subset of antennas, thereby reducing DL training and feedback overhead. For the far-filed antenna-domain CSI extrapolation, the challenge may result from the variation of the location of different antennas. It is much more challenging for the antenna-domain CSI extrapolation in near-filed, as the propagation paths may vary across antennas. In this case, the CSI extrapolation algorithm and the antenna selection scheme should be jointly designed.

*4) Multi-domain:* Most of the existing research on CSI extrapolation focus on the above single domain, i.e. time-domain, frequency-domain or antenna-domain. However, with the advent of 6G, joint CSI extrapolation in multi-domain illustrated in Fig. 2(d) becomes inevitable [51]. For example, joint frequency-antenna-domain CSI extrapolation is generally required in 6G, as the higher the spectrum exploited, the larger the MIMO systems needed to compensate the propagation loss and for directional beamforming. The challenge of joint CSI extrapolation in multi-domain arises from the fact that the channel correlation in a single domain is generally coupled, for example, the amplitude $\alpha_{i,j,l}(f,t)$, phase $\eta_{i,j,l}(f,t)$ and transmission delay $\tau_l(t)$ in Eq. 1 are affected by both the carrier frequency and location of the antennas. Moreover, in a V2X scenario where high-speed UEs are served using ultra-massive MIMO, joint time-antenna-domain CSI extrapolation is required to acquire CSI with low overhead. For a virtual reality (VR) scenario, where the UEs are served with ultra-massive MIMO using multiple spectrum bands for high throughput, joint frequency-antenna-domain CSI extrapolation is beneficial to minimize the overhead for CSI acquisition.

TABLE IV
Complexity Comparison of Different Models

| Model | Per-Layer Complexity |
|---|---|
| MLP | $\mathcal{O}(d_{MLP}^{out} d_{MLP}^{in})$ |
| RNN | $\mathcal{O}(T \times h \times (h + d_{RNN}^{in}))$ |
| LSTM | $\mathcal{O}(T \times 4 \times h \times (h + d_{LSTM}^{in}))$ |
| GRU | $\mathcal{O}(T \times 3 \times h \times (h + d_{GRU}^{in}))$ |
| CNN | $\mathcal{O}(H_{out} \times W_{out} \times C_{in} \times C_{out} \times H_k \times W_k)$ |
| GNN | $\mathcal{O}(K_s N_s F + N_s F^2)$ |
| Transformer | $\mathcal{O}(T^2 d_{Trans})$ |

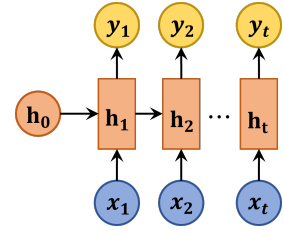

Fig. 4. Illustration of RNN.

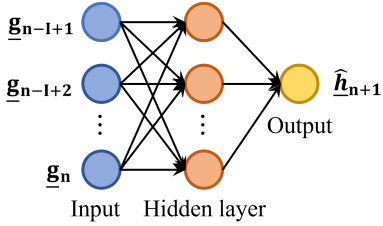

Fig. 3. Illustration of MLP.

A comparison of time-domain, frequency-domain, antenna-domain and multi-domain CSI extrapolation for 5G and 6G in terms of significance, applications & challenges are summarized in Table III.

### B. Fundamentals of AI

AI models plays a critical role in CSI extrapolation, and we will elaborate the applications of typical AI models for CSI extrapolation.

*1) MLP:* As illustrated in Fig. 3, MLP is a classic feedforward neural network architecture comprising an input layer, an output layer, and at least one hidden layer. Denote $d_{MLP}^{in}$ and $d_{MLP}^{out}$ as the number of input and output neurons, respectively. The computational complexity of an MLP layer is $\mathcal{O}(d_{MLP}^{out} d n_{MLP}^{in})$. It finds extensive application in multiple domains, including channel estimation and prediction tasks within wireless communication systems. Through multi-layer nonlinear transformations, the MLP can learn complex mapping relationships between inputs and outputs, rendering it suitable for regression problems such as CSI extrapolation. The structure of an MLP typically comprises an input layer, multiple fully connected hidden layers, and an output layer. Each hidden layer consists of numerous neurons, with each neuron performing a linearly weighted sum of its inputs before outputting through a nonlinear activation function. The core principle of the MLP lies in progressively extracting high-level features from input data through multi-layer nonlinear transformations, thereby approximating complex functional relationships. In channel prediction tasks, the MLP's input is typically a preprocessed received signal vector, with the output being the predicted future channel vector. Specifically, the MLP's input-output relationship can be expressed as [52]:

$$\hat{\underline{h}}_{n+1} = f_{\Pi}(\underline{g}_{n-I+1}, \cdots, \underline{g}_n), \qquad (10)$$

where, $\Pi$ denotes the parameter set of the MLP, $I$ represents the input order used to balance model complexity and predictive performance, and $g_n$ is the received signal vector preprocessed via LMMSE to enhance the model's robustness against noise. During the training phase of the MLP, the input comprises a preprocessed sequence of channel vectors $(\underline{g}_{n-I+1}, \cdots, \underline{g}_n)$, with the output being the channel vector $\hat{h}_{n+1}$ for the subsequent time step. To accommodate the real-valued neural network architecture, the complex input is typically decomposed into real and imaginary components. The hidden layer employs $L$ fully connected layers, each containing $f_l$ neurons. The output layer corresponds to the real and imaginary parts of the predicted channel vector, which are subsequently merged into a complex-form predicted channel vector via a reconstruction layer. ADAM is employed as the optimizer during training, with the loss function defined as the mean squared error between the predicted channel and the preprocessed channel:

$$C_{\text{loss}} = \frac{1}{N_{\text{train}}} \sum_{n=1}^{N_{\text{train}}} \left\| \hat{\underline{h}}_{n+1} - \underline{g}_{n+1} \right\|^2, \qquad (11)$$

where $N_{\text{train}}$ denotes the number of training samples. In CSI extrapolation tasks, the MLP learns temporal correlations within historical channel data to predict future channel states. Its advantage lies in the low computational complexity of the prediction phase once training is complete, making it suitable for communication scenarios demanding high real-time performance. However, MLP training demands substantial sample data and exhibits sensitivity to input order and network architecture. Appropriate configuration based on factors such as user mobility is necessary to balance performance and computational complexity.

*2) RNN-based models:* As illustrated in Fig. 4, recurrent neural networks constitute a specialised neural network architecture designed for processing sequential data, capable of capturing dynamic features within time series through internal states. Within the field of wireless communications, RNNs have become a vital tool for time-varying channel prediction owing to their ability to model temporal dependencies. The core of an RNN lies in its recurrent structure, which transmits information between different time steps through shared parameters and hidden states. At each time step, the RNN receives the current input and the hidden state from the previous time step, computing the current output and the updated hidden state. This recurrent computation leads to a per-layer time complexity of $\mathcal{O}(T \times h \times (h + d_{RNN}^{in}))$, where $T$ is the sequence length, $h$ is the hidden layer size, and $d_{RNN}^{in}$ is the input dimension. This mechanism enables RNNs to effectively process channel data exhibiting temporal correlation.

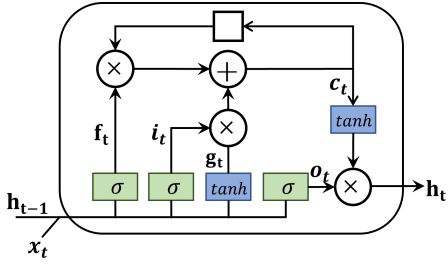

Fig. 5. Illustration of LSTM.

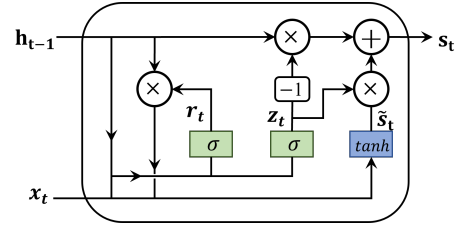

Fig. 6. Illustration of GRU.

In channel prediction tasks, the fundamental computational process of an RNN can be represented as follows [53]:

Hidden state updates as:

$$\mathbf{h}_t = \sigma(\mathbf{W}_{hh}\mathbf{h}_{t-1} + \mathbf{W}_{xh}\mathbf{x}_t + \mathbf{b}_h), \tag{12}$$

Output is calculated as:

$$\mathbf{y}_t = \mathbf{W}_{hy}\mathbf{h}_t + \mathbf{b}_y, \tag{13}$$

where, $h_t$ denotes the hidden state at the current time step, $x_t$ represents the current input, and $y_t$ signifies the current output. $W_{hh}$, $W_{xh}$, and $W_{hy}$ are the weight matrices, $b_h$ and $b_y$ are the bias vectors, and $\sigma$ is the activation function. Compared to traditional autoregressive models, RNN-based channel prediction methods better capture the nonlinear characteristics of channel variations, demonstrating strong adaptability in time-varying environments. The recurrent structure of RNNs naturally models the temporal correlation of channels, yielding more accurate predictions.

However, conventional RNNs face gradient vanishing or exploding issues when processing long sequences, limiting their performance in long-term channel prediction. Furthermore, the training process for RNNs is relatively complex, requiring careful hyperparameter tuning to ensure convergence.

As illustrated in Fig. 5, LSTM networks represent a specialized type of recurrent neural network that addresses the vanishing gradient problem encountered by traditional RNNs during training in long sequences by incorporating gating mechanisms and cellular states. LSTMs demonstrate exceptional performance in channel prediction tasks within the field of wireless communications, owing to their robust modeling capabilities for time-series data. The core innovation of LSTMs lies in their gating architecture and cellular state design. An LSTM unit comprises three key gating structures: the forget gate controls the retention of historical information, the input gate modulates the incorporation of new information, and the output gate determines the proportion of the current state's output. The cellular state serves as the carrier of long-term memory, sustaining information flow throughout the sequence processing. However, this sophisticated design comes with increased computational demands, resulting in a per-layer time complexity of $\mathcal{O}(T \times 4 \times h \times (h + d_{LSTM}^{in}))$, where $T$, $h$ and $d_{LSTM}^{in}$ denote the sequence length, hidden state size, and input dimension, respectively. In channel prediction, the primary computational steps of an LSTM include [54]: The forget gate determines which information to discard from the previous cellular state:

$$f_t = \sigma(W_f x_t + R_f h_{t-1} + b_f). \tag{14}$$

s Input gate control for new information addition is calculated as:

$$i_t = \sigma(W_i x_t + R_i h_{t-1} + b_i), \\ g_t = \tanh(W_g x_t + R_g h_{t-1} + b_g). \tag{15}$$

Cell Status Updates as:

$$c_t = f_t \odot c_{t-1} + i_t \odot g_t. \tag{16}$$

The output gate determines the final output as:

$$o_t = \sigma(W_o x_t + R_o h_{t-1} + b_o), \\ h_t = o_t \odot \tanh(c_t), \tag{17}$$

where, $W_f$, $W_i$, and $W_g$ denote the weight matrices from input $xt$ to the forget gate, input gate, and candidate state respectively. $R_i$, $R_g$, and $R_f$ represent the corresponding recurrent weight matrices and the recurrent weight matrix from hidden state $h_{t-1}$ to the forget gate. $W_o$, $R_o$, and $b_o$ are the output gate's weight and bias parameters, while $b_i$, $b_g$, and $b_f$ denote the bias vector of the forget gate. $\sigma$ is the sigmoid activation function, and $\odot$ denotes element-wise multiplication, enabling selective information updating.

As illustrated in Fig. 6, GRU represents an enhanced model based on recurrent neural network architecture, addressing the vanishing gradient problem encountered by traditional RNN when processing long sequences through the introduction of gating mechanisms. GRU demonstrates significant advantages in channel prediction tasks within the field of wireless communications, owing to their simplified structure and outstanding time series modeling capabilities.

The core innovation of the GRU lies in its gating mechanism design, which regulates information flow and memory state updates through two key gate structures: the update gate and the reset gate. Compared to LSTM, GRU merge long-term and short-term states into a single hidden state and reduce the number of parameters, enhancing computational efficiency while maintaining performance. This structural simplification yields a lower per-layer time complexity of $\mathcal{O}(T \times 3 \times h \times (h + d_{GRU}^{in}))$, where $T$, $h$ and $d_{GRU}^{in}$ denote the sequence length, hidden state size, and input dimension, respectively. This makes the GRU computationally more efficient than the LSTM.

In channel prediction tasks, the computational process of a GRU can be represented as [55]:

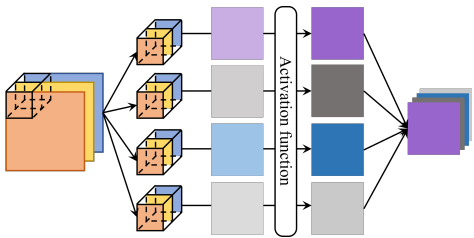

Fig. 7. Illustration of CNN.

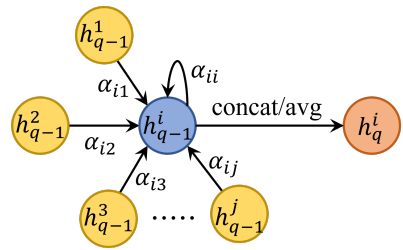

Fig. 8. Illustration of GNN.

Update door is calculated as:

$$\mathbf{z}_t = \sigma_g(\mathbf{W}_z\mathbf{x}_t + \mathbf{U}_z\mathbf{s}_{t-1} + \mathbf{b}_z). \qquad (18)$$

Reset door is calculated as:

$$\mathbf{r}_t = \sigma_g(\mathbf{W}_r\mathbf{x}_t + \mathbf{U}_r\mathbf{s}_{t-1} + \mathbf{b}_r). \qquad (19)$$

Candidate hidden state is generated as:

$$\tilde{s}_t = \sigma_h\left(\mathbf{W}_s\mathbf{x}_t + \mathbf{U}_s\left(\mathbf{r}_t \otimes \mathbf{s}_{t-1}\right) + \mathbf{b}_s\right). \qquad (20)$$

Final hidden status update as:

$$\mathbf{s}_t = (1 - \mathbf{z}_t) \otimes \mathbf{s}_{t-1} + \mathbf{z}_t \otimes \tilde{\mathbf{s}}_t, \qquad (21)$$

where, $\sigma_g$ denotes the sigmoid activation function, $\sigma_h$ denotes the hyperbolic tangent function, $W$ and $U$ are weight matrices, $b$ is the bias vector, and $\otimes$ denotes element-wise multiplication.

In channel prediction applications for multi-antenna systems, the GRU effectively captures temporal channel correlations through its gating mechanism. The update gate governs how much historical channel information is retained in the current state, while the reset gate determines how new channel observations are combined with the preceding state. This mechanism enables the GRU to adaptively balance long-term dependencies and short-term variations, maintaining stable prediction performance in rapidly changing channel environments.

However, the performance of GRU remains influenced by network depth and training data quality. In deep network architectures, careful hyperparameter tuning is required to avoid overfitting issues. Simultaneously, sufficient training data is crucial to fully realize the potential of GRU in channel prediction.

*3) CNN:* CNN is a type of feedforward neural network featuring convolutional operations and deep structures, widely used in image processing. As shown in Fig. 7, in a CNN, the convolution operation applies a set of shared weights (convolution kernels) to an input feature map to produce an output feature map. Mathematically, the value of the output feature map at position $(i, j, k)$ in the $h$-th layer can be expressed as:

$$y_{i,j,k}^{h+1} = g\left(\boldsymbol{w}_k^{h^T}\boldsymbol{x}_{i,j}^h + b_k^h\right), \qquad (22)$$

where $\boldsymbol{x}_{i,j}^h$ is the local receptive field of the input feature map at position $(i, j)$, $\boldsymbol{w}_k^h$ is the kernel for the $k$-th output feature map, $b_k^h$ is the bias term, and $y_{i,j,k}^{h+1}$ is the value of the output feature map at position $(i, j, k)$.

The dimensions of the output feature map are determined by the input feature map dimensions, kernel size, stride, and padding:

$$H_{out} = \left\lfloor \frac{H_{in} + 2P - H_K}{S} \right\rfloor + 1, \qquad (23)$$

$$W_{out} = \left\lfloor \frac{W_{in} + 2P - W_K}{S} \right\rfloor + 1, \qquad (24)$$

where $H_{in}$ and $W_{in}$ are the height and width of the input feature map, $H_K$ and $W_K$ are the height and width of the kernel, $P$ is the padding size, and $S$ is the stride. The computational complexity of a convolutional layer is primarily determined by the convolution operation itself, which can be expressed as $\mathcal{O}(H_{out} \times W_{out} \times C_{in} \times C_{out} \times H_k \times W_k)$, where $C_{in}$ and $C_{out}$ denote the number of input and output channels, respectively.

CNN excels at extracting local features through its convolutional layers, leveraging translation invariance to ensure efficient and accurate feature representation. Compared to FCNN, CNN reduces the number of parameters by sharing weights across spatial locations, making them computationally efficient for high-dimensional data [56]. However, due to their focus on local regions, CNN inherently struggles to model global relationships, particularly at a fine-grained level. This limitation has led to the integration of complementary techniques, such as attention mechanisms or global pooling layers, to enhance their ability to capture long-range dependencies.

*4) GNN:* As illustrated in Fig. 8, graph neural networks (GNNs) constitute a specialised deep learning architecture for processing graph-structured data, effectively capturing topological relationships and local dependencies between nodes. Within wireless communications, GNNs have emerged as a vital tool for CSI extrapolation tasks due to their inherent ability to model spatial correlations and local diffusion mechanisms. The core of GNNs lies in aggregating neighborhood information through graph-based message passing mechanisms to update node representations. In CSI extrapolation for fluidic antenna systems, each antenna port may be regarded as a node in the graph, with edges between nodes constructed based on spatial proximity relationships between ports. GNNs achieve gradual diffusion and reconstruction of channel features through multi-layer graph attention network modules. The computational process for each layer of the GAT module

is as follows:

$$e\left(\mathbf{h}_{q-1}^{i}, \mathbf{h}_{q-1}^{j}\right) = \mathbf{a}^{\top} \cdot \text{LReLU}\left(\mathbf{W}_e\left[\mathbf{h}_{q-1}^{i} \parallel \mathbf{h}_{q-1}^{j}\right]\right),$$

$$\alpha_{ij} = \frac{\exp\left(e\left(\mathbf{h}_{q-1}^{i}, \mathbf{h}_{q-1}^{j}\right)\right)}{\sum_{j' \in \mathcal{N}_i} \exp\left(e\left(\mathbf{h}_{q-1}^{i}, \mathbf{h}_{q-1}^{j'}\right)\right)},$$

$$\mathbf{h}_q^i = \sigma\left(\sum_{j \in \mathcal{N}_i} \alpha_{ij} \cdot \mathbf{W}_a \mathbf{h}_{q-1}^{j}\right), \tag{25}$$

Here, $h_{q-1}^i$ denotes the feature vector of the $i$-th port in the $q-1$-th layer, $a$, $W_e$, and $W_a$ represent learnable linear transformation matrices, LReLU is the nonlinear activation function, $\parallel$ indicates vector concatenation, and $\alpha_{ij}$ is the normalized attention weight reflecting the importance of neighboring node $j$ to central node $i$. Through multi-layer stacking, GNNs progressively fuse channel information between distant ports, enabling effective extrapolation for unknown port channels.

Within the AGMAE framework, the GNN serves as the decoder component, tasked with reconstructing the channel state for all unknown ports through a local diffusion mechanism [57]. This process operates upon the basis vector generated by the encoder and the known port CSI. Essentially, it involves learning the combination coefficients of the basis vector, thereby achieving reconstruction of the entire channel matrix while preserving spatial smoothness and local correlations. The advantage of GNN lies in its linear complexity with respect to the number of nodes, specifically

$$\mathcal{O}(K_s N_s F + N_s F^2), \tag{26}$$

where $K_s$ denotes the number of neighbors, $N_s$ represents the total number of ports, and $F$ signifies the feature dimension. This renders GNN particularly well-suited for large-scale CSI extrapolation tasks in high-resolution fluidic antenna systems.

However, the performance of GNNs is highly dependent on the quality of the constructed graph and the actual correlations between nodes. Should the graph structure fail to accurately reflect the genuine dependencies between channels, extrapolation accuracy may be constrained. Furthermore, GNNs exhibit sensitivity to the distribution of training data, and their generalisation capabilities within dynamic channel environments require further optimisation.

*5) Transformer:* As illustrated in Fig. 9, Transformers are groundbreaking neural network architectures built entirely on the attention mechanism, which have revolutionized AI across multiple domains. Their unified encoder-decoder structure and multimodal capabilities make them highly versatile and adaptable. Transformers are mainly composed of $N_e$ cascaded encoders and $N_d$ cascaded decoders. The core of the Transformer is the self-attention mechanism, which assigns dynamic weights to different parts of the input sequence, enabling the model to focus on the most relevant features. The self-attention mechanism results in a per-layer time complexity of $\mathcal{O}(T^2 d_{Trans})$, where $T$ is the sequence length and $d_{Trans}$ is the feature dimension. Mathematically, the self-attention output for position $i$ can be written as

$$\text{Attention}_i = \text{softmax}\left(\frac{(q_i W^Q)(K W^K)^{\top}}{\sqrt{d_k}}\right)(V W^V), \tag{27}$$

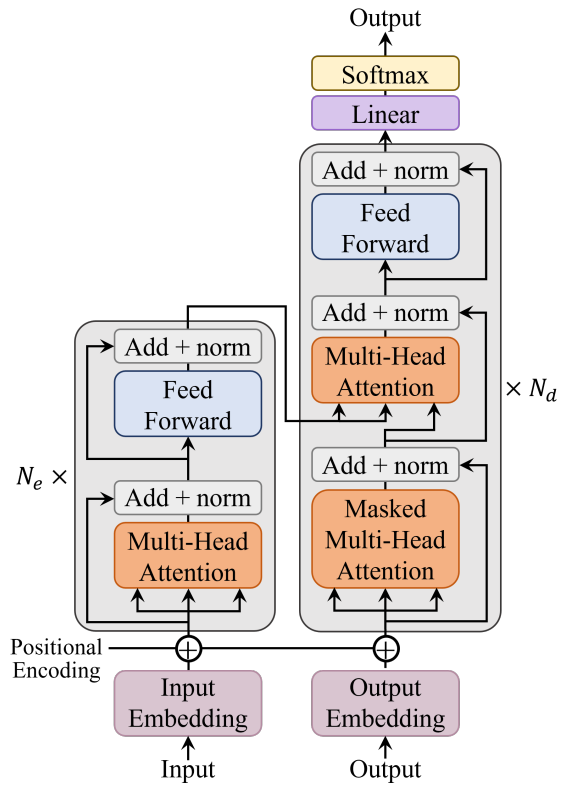

Fig. 9. Illustration of Transformer.

where $q_i$, $K$, and $V$ are the query, key, and value representations; $W^Q$, $W^K$, and $W^V$ are projection matrices; $d_k$ is the scaling factor; and softmax is the softmax function.

Transformers streamline feature processing with a unified architecture, but their $O(n^2)$ computational complexity for attention mechanisms imposes high resource requirements, particularly for long sequences. Moreover, the reliance of Transformers on large datasets for training can limit their efficiency on smaller datasets or simpler tasks, where models like CNNs may perform better.

Computational complexity of the above models are compared in Table IV.

## III. TIME-DOMAIN CSI EXTRAPOLATION

As illustrated in Fig. 2(a), time-domain CSI extrapolation acquires future CSI sequence using the historical CSI sequence, which is especially beneficial for high-speed communication scenarios, and the state-of-art time-domain CSI extrapolation schemes are reviewed comprehensively in this section. A comprehensive review of the state-of-art time-domain CSI extrapolation schemes are illustrated in Fig. 10.

### A. Model-Driven Approaches

Model-driven schemes rely on the temporal correlation of the channel, using mathematical models to describe channel variations over time. The core assumption is that the channel exhibits a degree of continuity, allowing future CSI to be derived from historical CSI through mathematical relationships.

| Method | Schemes | Principles | Strengths | Weaknesses |
|---|---|---|---|---|
| Model-based | AR-based | Future channel modeled as a linear combination of the historical channel with additive stochastic noise. | • Easy implementation;<br>• Computational efficient. | • Unfit for channel with strong channel aging effect. |
| | Parametric channel-based | Channel modeled mathematically regarding multipath components, Doppler frequency shift, etc., future channel calculated using the estimated parameters. | • Consider the physical properties of the channel, i.e., strong interpretability;<br>• High accuracy if model fits the channel. | • Limit to quasi-static channel with time-invariant or slowly varying channel parameters;<br>• Computation-intense for high accuracy parameter estimation. |
| AI-based | MLP | An MLP is a multilayer-perceptron model trained to fit the complex nonlinear dynamics of the channel in time domain. | • Enable complex nonlinear relationships. | • Dependent on large datasets;<br>• Difficult to train;<br>• Low parameter efficiency. |
| | RNN-series | Model the time-domain channel extrapolation as a sequence-to-sequence problem, process the input CSI sequence serially. | • Designed to process sequential data, effectively capture the time-series correlations inherent in communication channels;<br>• Variable sequence length handling. | • Error accumulation due to sequential extrapolation;<br>• Sequential extrapolation induced long-latency for large historical CSI sequence, unfit for high-speed communication. |
| | Transformer-based | Model the time-domain channel extrapolation as a sequence-to-sequence problem, process the input CSI sequence parallelly. | • Capture complex features and long-range temporal dependencies of the input historical CSI sequences using multi-head self-attention mechanisms;<br>• Parallel computation effectively mitigate channel aging in high-mobility scenarios. | • Computation-intense for training and inference;<br>• Positional encoding need improvements for time-series channel. |
| | Hybrid-AI schemes | Incorporating multiple AI models to capture the temporal correlation of historical CSI and extrapolate future CSI series. | • ConvLSTM enhances its spatio-temporal feature extraction abilities by substituting all fully-connected network in LSTM by CNN.<br>• CNN-Transformer enhances its ability to extract spatial information of XL-MIMO by adding a CNN at the beginning of the Transformer encoder. | • Weaknesses of multiple AI models;<br>• Difficult design to balance local and temporal features. |

Fig. 10. A comprehensive review of the state-of-art time-domain CSI extrapolation schemes.

This approach depends on pre-established models of channel dynamics and can be classified into the follows types:

*1) AR-based Approaches:* The AR model assumes that the current CSI is a linear combination of CSI values from previous time instances with additive stochastic noise. This is a classic time-series prediction method, characterized by simplicity and ease of implementation and various schemes have been proposed to capture the temporal correlations [52], [58]–[70]. [67] proposes a two-stage partial online optimization scheme for CSI extrapolation. Specifically, the first stage predicts the CSI of the next time slot using the CSI of the previous $L$ time slots and adapts to different mobile speeds by optimizing the AR model parameters online. In the second stage, the CSI of the unguided time slots is predicted by interpolation operation using the CSI predicted in the first stage and the actual CSI of the first $m$ time slots. With known temporal correlations, the Kalman filter (KF)-based predictor [52], [59], [66] and the Wiener filter-based predictor [60]–[62] are proposed. For example, [52] proposes a first-order polynomial-based extrapolation method to extrapolate time-varying channels. This extrapolation method uses the channel estimates from the last two time points to extrapolate the channel at the next time point. This extrapolation method is able to capture the time-varying characteristics of the channel better than using the outdated channel directly, thus improving the extrapolation performance. [62] proposes Wiener filtering for time-domain channel prediction. The scheme predicts the channel state at the current moment by using the pilot sequences sent by users in history.

The AR-based models assume explicit linear channel models, which is easy to understand and amenable to theoretical analysis. AR-based models are straightforward to implement, making them suitable for real-time applications and resource-constrained devices. The AR-based models show good performance for the CSI extrapolation of linear dynamic channel. However, AR-based models perform poorly in nonlinear or complex environments (e.g., scenarios with significant multipath effects or high mobility). Such drawback calls for sophisticated time-domain CSI extrapolation schemes.

*2) Parametric Channel-Based Approaches:* Parametric wireless channel model-based approach performs time-domain CSI extrapolation based on the quasi-static assumption, meaning that the key parameters, such as complex amplitude, delay and Doppler frequency shift of the wireless channel are time-variant or slowly varying in the extrapolation period [71]–[83]. Depending on the technique for parameter estimation, the parametric wireless channel model-based approaches include variational modal decomposition [73], maximum likelihood based method [74], [80], subspace based method, such as multiple signal classification (MUSIC) [75], [76], the compressed sensing based method [77], [78], estimation of signal parameters via rotational invariance techniques (ESPRIT) [75], [79].

Parametric channel-based approaches are based on channel models considering the physical characteristics which have explicit physical meanings with strong interpretability. Such approaches are especially effective if the channel model fits the channel scenarios. However, the performance of these approaches degrades dramatically if the channel is dynamic, i.e., the quasi-static assumption is invalid. In addition, it is generally computationally complex to estimate key parameters with high accuracy, limiting their applicability in computation-restricted scenarios with real-time requirements.

### B. AI-driven approaches

Unfortunately, due to the multi-path effect and the Doppler effect, practical channels usually evolve over time complicatedly, which makes the model-driven approaches difficult to match the actual channel. AI-based schemes employ machine learning or deep learning techniques and are promising to learn the complex patterns and long-term dependencies in channel variations from historical CSI data, making high-accuracy time-domain CSI extrapolation possible. Unlike model-driven schemes, they do not rely on predefined mathematical models but instead extract features and patterns through data-driven approaches. The state-of-art AI-driven time-domain CSI extrapolation approaches are reviewed comprehensively in the following section.

*1) MLP-based schemes:* Multilayer perceptron (MLP), utilizing a fully-connected architecture, is the first deep learning model proposed to extrapolate the future CSI sequence based on the learned channel characteristics from the historical CSI sequence [52]. However, as the input of MLP for time-domain extrapolation generally contains large channel information, it is challenging to train such models.

*2) Recurrent Neural Network (RNN) series-based Approaches:* Time-domain CSI extrapolation can be modeled as a sequence-to-sequence transformation from the historical CSIs to future CSIs, which could be enabled by RNN-based models [42], [52]–[55], [84]–[95]. In [84], a real-valued weight RNN-based channel prediction method is proposed and outperforms complex-valued weighted RNN models in terms of lower computational complexity and higher prediction accuracy. RNN-based time-domain CSI extrapolation models are also proposed in MIMO systems [42], [85], and shows better robustness to noise and interpolation errors than AR-based model with moderate computational complexity.

However, RNN-based models still show poor long-term dependencies of channel information and the evolution of RNN, i.e., LSTM [53], [54], [96], [97] and GRU [55], [95], [98], are proposed. Research in [91] shows that LSTM-based models learn the long-term and short-term dependencies of the time-series data much better than the baseline RNN, and have a better performance in the time-domain CSI extrapolation. In addition, LSTM-based models outperforms traditional Kalman and RNN model in LEO satellites systems [92]. [93] proposes a deep learning CSI extrapolation model based on LSTM networks, which is good at capturing correlations in time-series data, and thus can effectively predict future channel parameters and outperform the traditional AR integrated moving average and support vector regression (SVR) methods. [94] observed that the LSTM- and GRU-based models outperform RNN in terms of computational complexity and prediction accuracy in multi-antenna fading channels.

RNN series-based schemes are intuitive for time-domain CSI extrapolation owing to their capability of sequence-to-sequence transformation capturing sequential temporal dependencies. In addition, RNN series-based schemes are capable to process variable length of input sequence, making them flexible to coping with varying number of historical CSIs. However, RNN-based models perform CSI extrapolation sequentially, and the extrapolation error in the first few time steps accumulates in the following time steps. Also, the sequential extrapolation induces long-latency for large historical CSI sequence, which is undesirable for high-speed communication.

*3) Transformer-Based Approaches:* With the emerging 6G, the communication scenarios are becoming increasingly complex and complicated, conventional AI-based models struggle to achieve desirable performance. Due to the extraordinary long-range feature extraction capabilities, Transformer has been proposed for time-domain CSI extrapolation. The self-attention mechanisms is exploited to extract the deep and hidden relationships within CSI in time-domain [22], [38], [99], [100], [100]–[103]. [99] proposes a Transformer-based parallel CSI extrapolation scheme to solve the channel aging problem in mobile mmwave massive MIMO systems. Specifically, this paper transforms the CSI extrapolation problem into a parallel channel mapping problem, avoiding the error propagation problem in the traditional sequential extrapolation method. The proposed framework outperforms existing sequential CSI extrapolation methods (e.g., LSTM, RNN, etc.) in terms of NMSE performance and throughput performance in mobile scenarios with various mobility. [22] further improve the time-domain CSI extrapolation by removing the positional encoding. In time-domain CSI extrapolation, the temporal CSI series often lacks intrinsic semantics. The primary focus is on modeling the temporal variations in a continuous set of CSI values. Consequently, the sequence order in temporal CSI is vital for accurately understanding the temporal corrections of the CSI sequence. Adding positional encoding to the CSI sequence may lead to the loss of original sequence information or positional context, ultimately degrading extrapolation performance.

To summarize, Transformer-based approaches excel at capturing complex temporal features, making them especially effective for time-domain CSI extrapolation in complicated scenarios. Transformers are able to capture long-term temporal correction, which refers to the relationships between CSIs far apart in time step. Parallel computation in Transformer effectively mitigate channel aging in high-mobility scenarios. However, they require large datasets and significant computational resources for training. In additional, the inference of these models are also computation-intense, which is challenging to guarantee time-domain CSI extrapolation in real time. In addition, the permutation-invariance brought by the positional encoding does not perfectly suit to process time-domain CSI sequence.

*4) Hybrid-AI schemes:* Apart from the above single-model-based time-domain CSI extrapolation schemes, hybrid schemes that incorporates multiple AI models have been proposed. To enhance the performance of time-domain CSI extrapolation in complex scenarios, one of the major trends is exploiting the outstanding feature extraction ability of CNN and integrate it with other AI models to performance sequence-to-sequence transformation [104]–[110]. CNN is first integrated with RNN [109], and futher with LSTM [106]–[108]. Convolutional LSTM (ConvLSTM), where all the fully connected layer in LSTM are replaced by the CNN, is a classic model for sequence-to-sequence transformation, e.g., nowcasting precipitation [111], and has been tailored for time-domain CSI extrapolation. Specifically, CNN models in ConvLSTM enhance the spatio-temporal feature extraction abilities of conventional LSTM-based approaches, thus enhancing the performance of time-domain CSI extrapolation [112], [113]. In addition, CNN-Transformer enhances its ability to extract spatial information of XL-MIMO by adding a CNN at the beginning of the Transformer encoder [104].

Although hybrid schemes can exploit the strengths of multiple AI models, it is difficult to design such schemes. Taking ConVLSTM-based schemes as an example, for channels with varying coherence times (e.g., short in high-mobility scenarios), designing CNN kernels to capture time-localized patterns without distorting the sequence length for LSTM is tricky. In short coherence time scenarios, the CNN needs to focus on time-localized patterns (e.g., rapid phase shifts due to high Doppler), but aggressive downsampling or large kernels can blur these patterns. In long coherence time scenarios, the CNN must avoid overemphasizing short-term noise, while the LSTM needs a longer sequence to capture stable trends. In addition, the drawbacks in terms of sequential extrapolation of LSTM exist for ConVLSTM-based schemes, and needs to be resolved.

## IV. FREQUENCY-DOMAIN CSI EXTRAPOLATION

As illustrated in Fig. 2(b), frequency-domain CSI extrapolation acquires CSI in frequency point(s) of interest using the CSI in other frequency point(s), which is especially beneficial for FDD and multi-band communication scenarios, and the the state-of-art frequency-domain CSI extrapolation schemes are reviewed comprehensively in this section. A comprehensive comparisons between model-driven and AI-driven approaches for frequency-domain CSI extrapolation are summarized in Fig. 11.

### A. Model-Driven Approaches

Conventional approach for CSI extrapolation of FDD systems is the parametric wireless channel model-based approaches, which model the channel using physical properties, such as multipath components, to extrapolate CSI across frequencies [114]. By estimating parameters like path loss or delay spread, these methods reconstruct the channel response, often using tools like maximum likelihood estimation or subspace techniques [115], [116]. Research in [117] states that the extrapolation error is small when the frequency separation is less than half of the coherent bandwidth of the channel, but beyond that the error increases rapidly. [118] proposed a frequency-domain CSI extrapolation method based on the extrapolation matrix (EM), which greatly reduce the computational complexity compared to existing least squares (LS)- or minimum mean squared error (MMSE)-based methods. [119] proposes to estimate the DL channel covariance matrix by exploiting the frequency invariance of the angular scattering function of the UL channel to estimate the DL channel covariance from the UL channel parameters without additional DL channel training overhead. Unlike methods based on compressed sensing or dictionary learning, the method does not need to assume that the channel has discrete and sparse angular characteristics.

To further improve the CSI extrapolation accuracy, extrapolation methods based on high-resolution parameter estimation (HRPE) have been widely used in the frequency domain [120]–[122]. The HRPE method exploits the multipath structure of the channel to achieve a wider range of frequency-domain extrapolation, which provides an important reference for low-feedback overhead solutions in FDD massive MIMO systems. [123] points to the use of HRPE methods such as the space-alternating generalized expectation (SAGE) estimator to improve CSI extrapolation performance. The proposed HRPE-based extrapolation method can achieve a larger extrapolation range compared to the traditional LS and linear minimum mean square error (LMMSE) extrapolation methods. Compared to other extrapolation algorithms such as compressed sensing, the HRPE-based extrapolation algorithm performs better in terms of frequency domain extrapolation performance. [124] utilizes the SAGE algorithm to extract multipath channel parameters including amplitude, delay and angle information.

Parametric channel-based methods rely on predefined frequency response models, which are easy to understand and validate. They perform well if the model match the scenario. However, they struggle with model mismatches issue in complex multipath scenarios prevalent in 6G wideband systems, where it is challenging to model multi-path characteristics comprehensively. In addition, the high computational complexity of parameter estimation is another major concern for parametric channel-based methods.

### B. AI-Driven Approaches

*1) MLP:* MLP-based models are proposed for frequency-domain CSI extrapolation in FDD systems [125], [126]. MLP-based models are trained to learn the frequency correlation between subchannels, and are further used to extrapolate the unknown CSI of sub-channels. In [127], a channel predictor trained on the MLP architecture is proposed, which significantly reduces the training overhead and offers significant performance advantages over previous machine learning-based broadband channel prediction methods. To further improve the performance of CSI extrapolation, [128] proposed a DL CSI extrapolation method based on sparse complex-valued neural network, which learns the deterministic mapping relationship from UL to DL, and can directly project the DL CSI based on the UL CSI without the need of DL training and UL feedback. To accelerate the CSI extrapolation, [129] proposes a lightweight machine learning model to generate high-quality initial guesses, which are then improved using a highly efficient optimization framework that achieves extrapolation accuracies comparable to benchmark models an MLP-based method fast and efficient cross-band channel prediction.

| Method | Schemes | Principles | Strengths | Weaknesses |
|---|---|---|---|---|
| Model-based | Parametric channel-based | Channel modeled mathematically using physical properties, such as multipath components, etc., CSI across frequencies points are extrapolated using the estimated parameters. | • Consider the physical properties of the channel, i.e., strong interpretability; <br> • High accuracy if model fits the channel. | • Poor performance if the channel model mismatches with the scenarios,; <br> • Computation-intense for high accuracy parameter estimation. |
| AI-based | MLP-based | Trained to learn the frequency correlation between subchannels, and are further used to extrapolate the unknown CSI of sub-channels. | • Effective in learning complex mappings between frequency bands, potentially capturing non-linear frequency-selective characteristics. | • Low parameter-efficiency of MLP results in the large parameter volume of MLP and limits its scalability in multi-band systems. |
| | RNN-series | Modeling sequential patterns across frequency bands, potentially capturing correlations of CSIs across frequencies. | • Temporal information of CSI is utilized to improve the performance of frequency-domain channel extrapolation. | • CSIs across frequencies are not strictly sequential, reducing their effectiveness; <br> • Error accumulation due to sequential extrapolation. |
| | CNN | Perform frequency-domain channel extrapolation using the frequency-domain correlation capture by the kernels in CNN. | • Efficiently capturing local correlations among CSIs, showing high-performance for FDD systems. | • Limited long-range feature capturing, not suitable for frequency-domain channel extrapolation across wide frequency separation, such as multi-band systems. |
| | Transformer-based | Capture the correlation between CSI in frequency-domain using self-attention module, and perform sequence-to-sequence transformation. | • Ability to capture long-range feature across frequencies, promising for frequency-domain channel extrapolation in multi-band systems. | • Computation-intense for training and inference. |
| | Hybrid-AI schemes | A spatial attention module integrated into the CNN-based models to effectively extract internal connections among spatial features to improve CSI extrapolation. | • Focusing on the spatial features of the greatest importance, thereby improving performance of frequency-domain channel extrapolation; <br> • Easy integration between spatial attention module and CNN-based models. | • Relying on the assumption of input CSI spanning uniformly across frequencies. |

Fig. 11. A comprehensive review of the state-of-art frequency-domain CSI extrapolation schemes.

MLP can learn complex mappings between frequency bands, potentially capturing non-linear frequency-selective characteristics. However, the parameter-efficiency of MLP is low, resulting in the large parameter volume of MLP and limiting their scalability in multi-band systems. This implies that MLP requires large amounts of data to generalize across different frequency bands and scenarios, posing challenges for real-time deployment.

*2) RNN-series Approaches:* Temporal information of CSI is utilized to improve the performance of frequency-domain CSI extrapolation by using RNN, LSTM. In [130], an RNN-based frequency-domain extrapolation method for frequency-selective channels is proposed for wideband MIMO-OFDM systems. The model processes sequences of past CSI to forecast future states, accounting for fading dynamics. The RNN frequency-domain channel prediction method shows better performance on frequency-selective channels with greater flexibility and prediction capability than the conventional KF predictor. LSTMs extend RNNs by addressing the vanishing gradient problem, improving their training performance [131], [132]. In [133], an LSTM-based frequency-domain CSI extrapolation method for predicting path loss is presented. The proposed method improves the root mean squared error (RMSE) performance by more than 1 dB compared to the conventional method using the latest observation data, and the method maintains a high prediction accuracy in any frequency band including the high-frequency band. [134] presents an LSTM-based approach for predicting channel characteristics in realistic vehicular communication environments. The authors first designed a measurement campaign using an off-the-shelf On-Board Unit and a spectrum analyser to extract CSI from IEEE 802.11p frames, then collected real WAVE communication datasets from different vehicle-to-vehicle (V2V) driving scenarios, and finally constructed and evaluated a deep learning approach for predicting subcarrier-level CSI in these vehicular environments and frame-level received signal strength indicator (RSSI). [135] points out that AI-based methods, especially LSTM, RNN are considered as a promising technique for CSI prediction. The authors state that the prediction performance can be further improved by optimizing two parameters of LSTM-RNN, i.e., the number of input features and the number of hidden layers. The LSTM-RNN based methods have better prediction performance compared to the traditional estimation based methods, AR and ARMA modeling methods and second order statistical modeling methods. [136] proposes a deep learning-based frequency-domain CSI extrapolation method, SatCP, which avoids DL channel estimation by using DNNs to predict future DL CSI directly from observed UL CSI. Compared with the traditional method that requires DL channel estimation, SatCP method can effectively solve the challenge of obtaining effective DL CSI for LEO satellite massive MIMO systems.

RNN-series models is able to model sequential patterns across frequency bands, potentially capturing correlations of CSIs across frequencies. In addition, temporal information of CSI are utilized to improve the performance of frequency-domain CSI extrapolation. However, it is less intuitive for frequency-domain CSI, as frequency correlations are not strictly sequential, reducing their effectiveness compared to other models. In addition, RNN-based models perform CSI extrapolation sequentially, and the error accumulates for the wideband CSI extrapolation.

*3) CNN-Based Approaches:* CNN is adopted for frequency-domain CSI extrapolation attributed to its ability to capture the relationships across subcarriers. [137] proposes a CNN-based method to predict DL CSI using UL CSI to solve the problem of high feedback overhead of DL CSI in FDD large-scale MIMO systems. The method does not need to rely on channel sparsity assumptions, and the physical connection between the UL and DL bands is learnt by a neural network without any priori knowledge, using only the observed measurement data. [138] proposes a neural network method based on path gain coefficients, where the neural network is trained by extracting common path gain coefficients from UL and DL channels. The proposed method greatly reduces the input and output dimensions of the neural network, simplifies the training process, and maintains good performance even during high-speed movements.

CNN-based models exploit local frequency correlations, are similar to how they capture local patterns in images, making them suitable for handling frequency-selective fading. Thus, CNN-based models show good performance in FDD systems. However, due to the model design, CNN-based models struggle to capture the correlations of CSIs in wideband systems. It requires sophisticated design (e.g., filter sizes) to handle the specific structure of frequency-domain data, especially for wideband channels.

*4) Transformer-Based Approaches:* [139] proposed a Transformer-based CSI extrapolation for FDD systems. Unlike methods that first carry out UL channel estimation and then perform DL channel prediction in a step-by-step manner, [139] achieves an end-to-end framework that directly uses the UL pilot received at the BS as network inputs to achieve DL CSI. A hybrid feature extraction module based on encoder of Transformer with a non-causal attention mechanism is proposed to effectively capture the DL channel features, and the illustrate outstanding CSI extrapolation performance. [140] proposes a frequency-domain CSI extrapolation method based on Transformer. The network aims to estimate frequency-selective fading channels in RIS-assisted OFDM systems and employs the reflection patterns of some RIS reflective elements to reduce frequency conduction overhead. The proposed method can cope with different signal-to-noise ratios (SNRs) using a single training model, which reduces the offline training overhead and hardware cost, and can reduce the pilot overhead while coping with different noise levels, thus providing effective channel estimation for RIS-assisted OFDM systems.

Transformer-based approaches excel at capturing complex features in frequency domain, making them especially effective for frequency-domain CSI extrapolation in complicated scenarios. Transformers are able to capture CSI correction in frequency-domain, which refers to the relationships between CSIs far apart in frequency. This makes Transformer promising for CSI extrapolation for multi-band systems. However, they require large datasets and significant computational resources for training. Additionally, model training and inference demand significant computational resources, making them unsuitable for resource-constrained scenarios.

*5) Hybrid-AI schemes:* Apart from the above single-model-based time-domain CSI extrapolation schemes, hybrid schemes that incorporate multiple AI models have been proposed [141]. To effectively extract internal connections among spatial features by merging the channel information, a spatial attention module is integrated into the CNN-based models [142]. The spatial attention module is designed to be lightweight, incurring a negligible increase in computational complexity.

CSI in the frequency domain exhibits substantial variations in fading characteristics across subcarriers, particularly in channels with significant delay spread. The spatial attention mechanism enables the network to dynamically attend to the observed subcarriers that contribute most to the extrapolation of target subcarriers (e.g., those with similar path delays or higher power levels), thereby markedly improving the accuracy of both magnitude and phase estimation for the extrapolated subcarriers. In addition, by simply inserting one or two spatial attention modules into CNN-based architectures, significant performance gains can be achieved with minimal structural modifications and low implementation overhead. However, the spatial attention implicitly assumes that the observed subcarriers are uniformly or randomly distributed. When specific sparse sampling patterns are employed (e.g., nested sampling or coprime sampling), the attention mechanism may produce suboptimal weighting, leading to inferior performance compared with CNN architectures explicitly tailored to the given sampling scheme.

## V. Antenna-domain CSI extrapolation

As illustrated in Fig. 2(c), Antenna-domain CSI extrapolation acquires CSI in antennas of interest using the CSI of unobserved antennas, which is especially beneficial MIMO communication scenarios, and the the state-of-art antenna-domain CSI extrapolation schemes are reviewed comprehensively in this section. A comprehensive comparisons between model-driven and AI-driven approaches for antenna-domain CSI extrapolation are summarized in Fig. 12.

### A. Model-Driven Approaches

Model-driven methods typically leverage the physical characteristics of the channel (e.g., spatial correlation or angle of arrival) to predict the CSI of unobserved antennas within the same panel. This approach is particularly significant in massive MIMO systems, as it reduces training and feedback overhead. Model-driven schemes exploit the physical properties of the channel for extrapolation, relying primarily on two key attributes:

- Spatial Consistency Property: The channel responses between adjacent antennas are typically highly correlated, enabling the prediction of unmeasured antenna CSI from known antenna CSI. This property is crucial in millimeter-wave massive MIMO, especially under non-stationary channel conditions [143].

| Method | Schemes | Principles | Strengths | Weaknesses |
|--------|---------|-----------|-----------|------------|
| Model-based | Interpolation Method | Assuming spatial continuity or smoothness of CSIs in antenna-domain. | • Easy implementation;<br>• High performance spatially smooth channels. | • Unfit for dynamic channel environments;<br>• Limit to strong antenna correlation. |
| | Parametric channel-based | Channel modeled mathematically with multipath components regarding antenna array geometry, such as the amplitude, delay, AOA, etc., CSI of interest calculated using the estimated parameters. | • Consider the physical properties of the channel, i.e., strong interpretability;<br>• High accuracy if model fits the channel. | • Poor performance in multipath-rich scenarios, where the characteristics of all paths are challenging to acquire;<br>• Limit to strong spatial correlation, where the spacing between adjacent antennas is desirable. |
| AI-based | MLP-based | Learning a nonlinear mapping from CSI of partial antenna to the CSI of the entire antenna array. | • Simple to implement due to the simple structure of MLP. | • Difficult to capture spatial structures of CSI in antenna domain;<br>• The parameter-efficiency of MLP is low, resulting in the large parameter volume and limiting their scalability in ultra-massive MIMO. |
| | CNN-based | CNNs treat CSI in antenna-domain as a two-dimensional image and use convolutional layers to capture local correlations of CSIs between antennas. | • Efficiently capturing local spatial correlations for grid-like antenna arrays;<br>• Strong spatial-feature capturing capabilities. | • Spatial-feature capturing restricted to local due to limited receptive fields, not suitable for ultra-massive MIMO;<br>• Less effective for non-uniform arrays, such as fluid antenna systems. |
| | GNN-Based | GNNs model the antenna array as a graph, with antennas as nodes and spatial relationships as edges. | • Adapt to various antenna layouts makes them particularly appealing for antenna-domain channel extrapolation with irregular antenna layouts;<br>• Capturing long-range dependencies for large-spacing arrays. | • Difficult to construct the graph of the antennas, errors or oversimplifications in the graph can compromise extrapolation accuracy;<br>• Computational cost scales dramatically with the number of antennas. |
| | Transformer-based | The known CSI and CSI of interest in antenna domain transformed into sequence via positional encoding, sequence-to-sequence performance by Transformer. | • Learns long-range dependencies;<br>• High flexibility for dynamic channels and extrapolation;<br>• 2D positional encoding effective to model the relative position of the antennas. | • Computation-intense for training and inference. |
| | Hybrid-AI schemes | Incorporating AI model and mathematical model to effectively capture the correlation of CSI In antenna domain. | • Ordinary differential equation (ODE) utilized to describe the latent relation between different data layers, speeding the convergence and learning performance of CNN-based schemes;<br>• The antenna-domain CSI exhibits approximately linear phase variation along the array axis, which can be approximated by a first-order or second-order ODE. | • High computational complexity of solving ODE, unsuitable for latency-sensitive scenarios;<br>• ODE approximation invalid for ultra-massive systems, performance tends to degrade dramatically. |

Fig. 12. A comprehensive review of the state-of-art antenna-domain CSI extrapolation schemes.

- Angular Domain Sparsity: The channel is often sparse in the angular domain, meaning only a few multipath components' angles (e.g., angle of arrival, AoA) dominate the channel response, allowing reconstruction of the entire antenna array's CSI using these parameters. Studies support this in FDD multi-user massive MIMO systems, utilizing angular domain sparsity for compressed sensing estimation [144].

*1) Interpolation-based Approaches:* Interpolation-based methods assume spatial continuity or smoothness in the channel, estimating unmeasured antenna CSI via mathematical interpolation of known CSI, independent of specific physical parameters [145]. Linear interpolation performs linear interpolation between adjacent antenna CSI for antenna-domain CSI extrapolation, which is simple but with limited accuracy. Spline interpolation uses cubic spline functions to fit known CSI, offering smoother estimates for gradually varying channels. Kriging interpolation is a statistical method employing covariance models (e.g., Gaussian processes) for optimal unbiased estimation to enhance CSI extrapolation accuracy [146]. Spatial covariance matrix of the channel is proposed to be utilized to extrapolate CSI, which shows outstanding performance in scenarios with stable statistical properties, such as indoor environments or slow-fading channels [147].

Interpolation methods are initiative to understand and easy to implement. These models are computational-efficient and effective in spatially smooth channels. However, interpolation methods are unable to perform high-accuracy antenna-domain CSI extrapolation if CSIs for different antennas are not smoothly varying, such as the ultra-massive MIMO systems.

*2) Parametric Channel-Based Approaches:* Parametric wireless channel model-based schemes are utilized for antenna-domain CSI extrapolation, which extract multipath parameters (e.g., amplitude, delay, angle of arrival (AoA)) from physical channel models to extrapolate CSI for unmeasured antennas [148]. By utilizing array geometry and spatial correlation, a vector space signature (VSS) model is proposed to extract multipath parameters for CSI extrapolation [120]. AoA is estimated and combined with array response vectors for CSI extrapolation. To further enhance the CSI extrapolation accuracy, a high-resolution parameter estimation is employed algorithms like SAGE to decompose multipath components from received signals, which is effective for wideband MIMO systems [117]. [149] presents a dual-polarization CSI extrapolation methodology that initially estimates polarization-independent parameters using numerous pilots in one polarization direction before re-estimating them in the other direction, thus reducing pilot overhead. For millimeter-wave or ultra-massive MIMO systems with evident angular-domain sparsity, compressive sensing (CS) is proposed to extrapolate full CSI from partial measurements [81].

Parametric channel-based approaches are based on explicit physical models, depending on key parameters including amplitude, delay, AOA, etc. These methods are easy to understand and validate and are especially effective if the channel model fits the channel scenarios. Similar to the parametric channel-based approaches for CSI extrapolation, it is generally computational complex to estimate the key parameters with high accuracy. Additionally, in real-world environments, channels may exhibit spatial non-stationarity due to blockages or multipath effects, causing model assumptions to fail. For multipath-rich scenarios, models may fail to accurately capture all path characteristics, and the performance of the parametric channel-based approaches degrade dramatically. The performance of these approaches degrade dramatically if the channel is dynamic, i.e., the quasi-static assumption is invalid. When the spacing between adjacent antennas increases, spatial correlation weakens, leading to significant extrapolation errors.

### B. AI-driven approaches

With the involution of MIMO technologies towards ultra-massive, movable and programmable, explicit physical models fail to capture the characteristics of emerging MIMO channels. To this end, AI-driven approaches use machine learning or deep learning to learn spatial relationships between antennas using known CSI and extrapolate the unknown CSI in antenna-domain. This method avoids reliance on explicit physical models, instead using data-driven approaches to capture complex channel properties. The state-of-art AI-driven time-domain CSI extrapolation approaches are reviewed comprehensively in the following section.

*1) MLP-Based Approach:* MLPs are multi-layer fully connected networks that learn a nonlinear mapping from a vector of partial antenna CSI (typically flattened into real or complex tensors) to full array CSI. Training employs supervised learning with loss functions like mean squared error (MSE) or normalized MSE (NMSE), optimizing weights via backpropagation [21], [33], [150]. For example, [150] describes a method based on a ray model for extrapolating spatial channels, which assumes that differences in antenna positions can be modeled through phase shifts while keeping power, AoA, and AoD constant. This method outperforms static channel estimation by increasing the extrapolation distance. [151] presents a DNN-based CSI extrapolation method suitable for pattern reconfigurable massive MIMO systems, performing channel estimations with minimal overhead through grouping antennas by radiation patterns.

The structure of MLP is simple, making them easy to implement. By changing the input and output dimensions, MLPs are flexible to carry out CSI extrapolation with varying number of antenna elements. However, the structure of MLP makes it difficult to capture spatial structures of CSI in antenna domain. In addition, the parameter-efficiency of MLP is low, resulting in the large parameter volume of MLP and limiting their scalability in ultra-massive MIMO. Such parameter-inefficiency makes MLP trained with excessive CSI data in antenna domain to approximate the antenna domain correlations.

*2) CNN-Based Approaches:* CNNs treat CSI as a two-dimensional image and use convolutional layers to capture local correlations between antennas, extrapolating missing CSI portions [21], [152], [153]. [154] proposes a method for antenna-domain channel prediction using CNN to predict the statistical properties of the channel. The proposed method suggests better data acquisition rules and has a significant advantage in prediction accuracy over previous methods. [21] proposes a deep learning-based scheme for achieving antenna-domain CSI extrapolation, enhancing system efficiency and adaptability compared to traditional methods. [155] introduces FadeNet for large-scale channel fading prediction, demonstrating accuracy and speed by learning directly from data without manual tuning. [156] implements two CNN-based networks, IENet and CENet, for joint training to optimize performance in reducing intra-group interference in RIS systems.

CNN-based approaches are naturally suited for grid-like antenna arrays, capturing spatial correlations effectively through convolutional filters. However, they may struggle with irregular antenna configurations, such as fluid antenna systems, or non-stationary spatial characteristics, requiring modifications like graph-based approaches. In addition, as CNNs are less effective for capturing long-range dependencies, which refer to the relationships between CSIs far apart in antenna elements. Such weakness restricts their performance for ultra-massive MIMO.

*3) Graph Neural Networks (GNNs)-Based Approaches:* GNNs modeling the antenna panel as a graph structure, GNNs learn dependencies between antennas, making them particularly suitable for irregular antenna arrays. [57] proposes an asymmetric graph masked autoencoder (AGMAE) architecture for antenna-domain CSI extrapolation. Specifically, the Transformer attention mechanism is used in the encoder part and the graph attention network is used in the decoder part, and the whole architecture is asymmetrically designed to balance the computational complexity. The method supports different numbers of CSI inputs and can be extrapolated to different sizes of arrays by training only one fixed-size array, which provides good flexibility and generalization ability. Compared with traditional extrapolation methods such as compressed sensing and linear interpolation, the AGMAE method has obvious advantages in terms of flexibility, computational complexity and generalization ability.

GNNs offer a promising approach for antenna-domain CSI extrapolation by leveraging spatial relationships in a graph-based framework. Their ability to model complex dependencies and adapt to various antenna layouts makes them particularly appealing for antenna-domain CSI extrapolation with irregular antenna layouts or large-spacing arrays. However, their success hinges on overcoming challenges such as graph design, computational demands, and data availability. The effectiveness of GNNs heavily depends on the graph accurately reflecting the true relationships between antennas. Errors or oversimplifications in the graph can compromise extrapolation accuracy. Careful implementation and potential integration with domain knowledge or other techniques may be necessary to fully realize their potential in this application. GNNs involve message passing between connected nodes, which can become computationally expensive, especially for large antenna arrays, despite their scalability in principle.

*4) Transformer-Based Approaches:* Currently, Transformer also has some applications in antenna-domain CSI extrapolation [23], [157], [158]. The authors of [30] attempt to use machine learning algorithms such as linear regression and SVR for antenna-domain CSI extrapolation, and although some results were achieved, the traditional machine learning methods are difficult to adapt to large-scale antenna arrays due to their obvious spatial non-stationarity. To this end, the authors propose a Channel Transformer (CT) model based on a self-attentive mechanism, which establishes the correlation between different antenna positions, and the position embedding preserves the positional relationship between antennas, thus solving the spatial non-smoothness problem. Compared with methods such as CNN and RNN, CT can better handle the CSI information of long-distance antennas in ultra-large-scale MIMO systems, by utilizing the self-attention mechanism. In addition, through the transfer learning (TL) strategy, the CT can train adaptive AI models under different propagation environments to improve the extrapolation accuracy. [159] utilizes a reference-based variational auto-encoder to estimate full antenna domain channels in hybrid RIS-assisted mmWave systems, improving channel estimation performance. [158] presents a bidirectional encoder representations from Transformers (BERT) and masks language model-based spatial CSI extrapolation method that efficiently infers complete CSI from partial observations without additional pilot inputs, improving on traditional CSIs.

Transformer-based approaches can learn to generate channel states for unseen antenna configurations, potentially adapting to new spatial setups. However, high complexity and data requirements, and may not capture fine-grained spatial correlations without sufficient training, posing challenges for real-time applications. Model training and inference demand significant computational resources, making them unsuitable for resource-constrained scenarios.

*5) Hybrid-AI schemes:* To improve the feature extraction of

CNN, ordinary differential equation (ODE) [160] is applied in [161] for CSI extrapolation methods in reconfigurable intelligent surface (RIS)-assisted communication systems to extrapolate the full CSIs from the partial ones. The proposed ODE-CNN structure introduces cross-layer connectivity and linear computation for faster convergence and better performance than cascaded CNN methods. However, Neural ODEs require numerical integration of the underlying ODE system, typically employing adaptive or fixed-step solvers such as high-order Runge-Kutta methods or the Dormand-Prince method integrator [162], [163]. This increased computational overhead can constitute a critical bottleneck in ultra-low-latency applications, such as URLLC scenarios in 5G-Advanced and 6G systems [164]. Furthermore, antenna-domain CSI under far-field and line-of-sight (or single-dominant-path) conditions exhibits approximately linear phase progression along the array manifold, which can be effectively modeled by low-order (first- or second-order) continuous-time dynamics. However, in near-field regimes or rich-scattering environments with significant multipath components, the channel's spatial evolution deviates substantially from simple linear ODE assumptions. Consequently, the extrapolation accuracy of pure Neural ODE-based methods tends to degrade dramatically in such complex propagation conditions [165].

## VI. MULTI-DOMAIN CSI EXTRAPOLATION

As illustrated in Fig. 2(d), multi-domain CSI extrapolation can be regarded as the integration of time, frequency and antenna-domain extrapolation. [34] points out that although multi-domain joint CSI extrapolation can provide higher accuracy channel estimation, enhanced robustness, and reduced overhead compared to single-domain CSI extrapolation, its complexity is also higher than that of single-domain CSI extrapolation, with multi-domain data being more difficult to synchronize and calibrate, and a higher resolution of a prior information being required. Currently the existing research on multi-domain CSI extrapolation can be classified as model-driven schemes and AI-driven schemes.

### A. Model-Driven Schemes

Similar to the CSI extrapolation discussed in the previous sections in a single domain, model-driven schemes mainly focus on parametric channel-based approaches. The existing mainly exploited the multi-path channel model, similar to the Saleh-Valenzuela channel model [22], [166], where the key parameters are the number of paths, delay, AoA, DoA, Doppler shift and complex gain of each path, as a start point and developed correlations in each domain [122], [167].

[34] proposes a two-stage time-frequency-antenna-domain CSI extrapolation scheme in FDD MIMO systems. This work assumes a linear phase shift across subcarriers and antennas to model frequency-domain and antenna domain correlations, respectively. By assuming low-speed movement of UEs, the key parameters are time-invariant within each frame, i.e., parameter estimation period. The channel parameters of the next frame is summation of the channel parameters and

| Schemes | Principles | Strengths | Weaknesses |
|---------|-----------|-----------|------------|
| MLP-based | Integrating with assumption on wireless channel to capture effective correlations between CSI in multi-domain. | • Easy implementation to learn the mapping between the known CSI to the CSI of interest. | • Low parameter-efficiency of MLP results in the large parameter volume of MLP;<br>• Poor performance if the assumption on wireless channel mismatch the scenario of interest. |
| CNN | 3D convolutional kernels in CNN are effective to capture the correlation of the CSI in the temporal, antenna, and frequency domain. | • Excel at extracting coupled CSI correlations in multi-domain, making them well-suited for multi-domain CSI extrapolation tasks. | • Limited to local feature capturing, which restricts their applicability in joint high-speed, EL-MIMO and wideband systems, etc. |
| Transformer-based | Capture the correlation between CSI in multi-domain using self-attention modules and multi-dimensional positioning coding. | • Scaling-law utilized for effective multi-domain CSI correlation extraction. | • Computation-intense for training and inference;<br>• The efficiency of the baseline Transformer architecture to capture the joint correlation of CSI in multi-domains is in doubt. |
| Hybrid-AI schemes | Incorporating multiple AI models for effective CSI channel extrapolation. | • GAN used to generate synthetic channel data to enrich the training dataset, thereby enhancing the performance of CSI extrapolation module. | • Difficult to generate synthetic channel samples with specific scenario attributes, thereby limiting the generalization of the CSI extrapolation model in multi-domain. |

Fig. 13. A comprehensive review of the state-of-art AI-driven multi-domain CSI extrapolation schemes.

stochastic terms. Based on the above correlations, key parameters are estimated for multi-domain CSI extrapolation. [168] presents a parameterization-based prediction method for polarized narrowband MIMO based on 3GPP/WINNER II SCM model. The method utilities information in the time, antenna and polarization domains to jointly estimate the AoA, AoD, Doppler shift and complex polarization weights of multipath signals via multidimensional ESPRIT. The method is able to estimate the channel parameters more accurately than the one-dimensional ESPRIT method, which utilities only time-domain information, and therefore exhibits a lower NMSE over all prediction time scales. [169] presents a channel prediction method based on wavefront transform matrix pencil. The method designs a matrix that transforms a spherical wavefront into a new wavefront that is closer to a plane wave, thus mitigating the performance loss caused by near-field effects in the moving scenario of an extra-large antenna array. [170] proposes CSI extrapolation method based on the surface roughness parameter of a calibrated scatterer. The method achieves the extrapolation of the channel correlation matrix between the frequency and antenna domain and shows good performance. Extrapolation from the frequency domain CCM to the antenna domain CCM can reduce the prediction error of the variable part by about 50%. Extrapolation from the antenna domain CCM to the frequency domain CCM also reduces the prediction error of the variable part by about 50 per cent, which is comparable to the results obtained using frequency domain CCM extrapolation.

For multi-domain CSI extrapolation, parametric channel-based approaches are based on explicit physical models, either derived based on certain assumptions or commonly used channel models. Similar to CSI extrapolation approaches for a single domain as discussed in the previous sections, key parameters including number of paths, complex gain, delay, AOA, etc., are estimated in the first stage and utilized for CSI extrapolation. These methods are easy to understand and validate and are especially effective if the channel model fits the channel scenarios. However, the commonly used algo-rithms for high-accuracy parameter estimations are generally computational complex, such as MUSIC, ESPRIT, or evolved versions. In addition, it is challenging to model the channel comprehensively, for example, it is difficult to accurately capture all path characteristics in multipath-rich scenarios, and the performance of the parametric channel-based approaches will degrade. In addition, the existing research model the correlations between time, frequency and antenna-domain separately. However, the coupling of channel correlation across different domains are reported in the existing research, for example, the channel correlations in time domain and frequency domain are coupled in high-speed scenarios [171].

### B. AI-driven approaches

With the advent of 6G, joint CSI extrapolation in multi-domain becomes inevitable and is much complicated than CSI extrapolation in a single domain. To this end, sophisticated AI-driven approaches have been developed and the state-of-art schemes are reviewed comprehensively in this section. A comprehensive comparisons between AI-driven approaches for multi-domain CSI extrapolation are summarized in Fig. 13.

*1) MLP-Based Approach:* MLP has been applied in multi-domain CSI extrapolation. [35] proposes a channel mapping method based on interleaving learning, which designs a complex-domain MLP mixer model that learns the channel characteristics in antenna and frequency domain separately, and then captures the correlation between the two domains through interleaving learning. This design greatly reduces the learning burden and exhibits high efficiency in channel mapping performance. Utilizing the high angle-delay resolution of wideband massive MIMO systems, [36] is proposed to achieve multi-domain CSI extrapolation by exploiting the channel characteristics in angle-delay domain. The proposed supervised learning method based on deep learning, which uses a complex-valued neural network (CVNN) to predict channel elements in the angular-delay-domain.

The MLP-based approaches are easy to implement due to the simple structure of MLP. However, it is challenging for MLP to capture the coupled channel correlations in multi-domain effectively. To this end, existing research proposed sophisticated MLP-based network based on the prior assumptions of specific physical channel characteristics. This approach can be regarded as the combination of parametric channel-based approaches and MLP. Thus, intuitively, the performance of the MLP-based approaches depends on whether the assumptions of channel characteristics match the scenario of interest.

*2) CNN-based Approaches:* CNN has been proposed for multi-domain CSI extrapolation due to its strong feature extraction abilities. In [172], the authors propose a three-dimensional (3D) CNN-based deep learning framework for extrapolating future DL CSI, which utilizes 3D convolutional kernels to capture the correlation of the channel in the temporal, spatial, and frequency dimensions, and thus accurately extrapolates the CSI of interest. [37] proposes a knowledge-driven and data-driven spatial-frequency network. The method exploits an LS estimator for rough channel estimation and combines spatial-frequency CSI extrapolation to reduce the pilot overhead in both the spatial and the frequency domains. The authors propose a sub-element extrapolation module based on an attention mechanism and an asymptotic extrapolation architecture to improve the accuracy of spatial-frequency CSI extrapolation. [173] proposes a CSI extrapolation method based on deep residual U-shaped network for estimating the channel characteristics of RIS cascaded OFDM systems. To further improve the extrapolation performance, noise variance is adopted as an additional input to improve the performance of channel estimation at different SNRs, while using hopping connections to enhance the learning capability of the network and a single training model for different SNRs, which significantly reduces the amount of training data and training time. [174] proposes a deep learning-based scheme that combines spatial extrapolation and temporal prediction to efficiently estimate fast-changing cascading channels. The proposed spatial-temporal convolutional network, which can process a large amount of spatial-temporal data in parallel without the need to store a large amount of hidden states, thus providing high performance and efficient computation. To further improve the efficiency of the deep learning model, the authors implemented structured probabilistic pruning, which outperforms uniform and random pruning for extrapolation.

CNNs excel at extracting coupled channel correlations in multi-domain, making them well-suited for multi-domain CSI extrapolation tasks. However, as the CNNs focuses on local feature extraction, similar to the CNN-based approaches for single domain CSI extrapolation, only the local channel correlations can be coupled, which restricts their applicability of in joint high-speed, EL-MIMO and wideband systems. In addition, CNN-based approaches are computation-intensive, which could pose challenges for real-time implementation or deployment on resource-constrained devices.

*3) Transformer-Based Approach:* As mobile networks evolve towards complexity, generalization of deep learning-based CSI extrapolation models becomes a huge challenge. Emerging Transformer is expected to solve the CSI extrapolation problem in complex communication scenarios [51].

[139] proposes a Transformer-based end-to-end DL channel prediction neural network, E2ENet. It predicts the DL channel directly using the UL pilot signals and avoids the accumulation of UL channel estimation errors, thus improving the accuracy of the DL channel prediction. E2ENet is designed with a hybrid feature extraction module (HFEM) that simultaneously models the modeling of temporal, spatial and frequency features to better capture the relationship between the UL pilot frequency and the DL channel. The proposed E2ENet performs significantly better than the conventional two-step method and also predicts both UL and DL channels simultaneously with reduced storage overhead. [175] proposes a large language model-based channel prediction method, LLM4CP, to predict the future DL CSI sequences of a MIMO-OFDM system by fine-tuning the pre-trained generative pre-trained Transformer (GPT)-2 model, which can be simultaneously applied to both TDD and FDD systems. Compared with existing model-driven and deep learning methods, the NMSE of LLM4CP significantly outperforms that of other benchmark methods under different user rates and exhibits higher prediction accuracy. And LLM4CP is more robust to CSI noise and outperforms other methods in terms of NMSE performance. In the no-sample learning scenario (using only 10% of the training data), the advantage of LLM4CP is even more obvious, reflecting the strong no-sample learning capability. In addition, LLM4CP also outperforms other methods in terms of communication performance metrics such as SE and bit error rate (BER).

GAI approaches, such as those using generative models or large language models, are designed to handle complex and diverse communication scenarios, offering superior generalization. GAI models, particularly those involving large language models or intricate generative architectures, demand significant computational resources, increasing training and deployment costs. Although some GAI methods reduce pilot data needs, they may still require substantial datasets for effective training, especially across diverse scenarios. GAI models can be less interpretable than simpler approaches, making it difficult to understand or debug their predictions, which may hinder practical deployment.

*4) Hybrid-AI schemes:* Due to the complexity and challenging of multi-domain CSI extrapolation, hybrid-AI schemes have attracted wide attention to integrate the abilities of various AI models. One of the most porpoising approach is generating channel data to enhance the performance of multi-domain CSI extrapolation. In [182], a GAN-LSTM framework is proposed to extrapolate CSI for 6G. The framework uses GAN to generate synthetic channel data to enrich the training dataset, while using LSTM for sequence prediction to predict future channel characteristics. By combining GAN and LSTM, the challenge of insufficient datasets is effectively addressed, the quality and diversity of the required channel data are greatly improved, and the efficiency of parameter generation outperforms that of traditional channel modeling approaches. However, GAN cannot be easily controlled to generate data with specific scenario attributes, restricting the variability of the synthetic channel samples, which may restricts the performance of CSI extrapolation model in different or multiple scenarios with high variations of channel characteristics.

TABLE V
An overview of open-sourced wireless channel dataset applicable for the research of CSI extrapolation. TR and M are short for ray tracing and measurement.

| Name (collection/ generation) | Scenarios | Frequency band (bandwidth) | Applicability | Limitations | Strengths |
|---|---|---|---|---|---|
| Industrial radio (M) [176] | Industrial environment | 3.7-3.8 GHz (80 MHz) | Time-domain | Low mobility unsuitable for high-mobilty sceanrio. | High measurement frequency of 1kHz for fine-grained time-domain CSI extrapolation |
| DICHASUS (M) [177] | Industrial environment | 1.272 GHz (50.056 MHz) | Time-frequency-antenna-domain | Large measruement interval of 48ms MIMO of 32 antennas | CSI of 1024 OFDM subcarriers in band frequency-domain CSI extrapolation. |
| Wireless Intelligence (M)[I] | Outdoor campus | 2.565 GHz (20 MHz) | Timer-frequency-antenna domain | MIMO of 4 TX and 2 Rx Static measurement at each site | High measurement frequency of 1kHz for fine-grained time-domain CSI extrapolation |
| WARI-D (RT) [178] | Outdoor | 2.6, 6, 28, 60, 100 GHz | Frequency-antenna -domain | Static channel | 64 and 384 antennas @ 6 and 28 GHz Covering sub-6, mmWave spectrum bands |
| DeepMIMO (RT) [179] | Outdoor, indoor | 3.4, 3.5, 60 GHz | Frequency-antenna -domain | Static channel | Covering sub-6, mmWave spectrum bands Configurable antenna number beneficial for ferquency-domain CSI extrapolation |
| DataAI-6G (RT) [50] | Outdoor, indoor | 3.5, 6.5, 28 GHz | Time-frequency-antenna-domain | Small size of MIMO, UE moves towards a single direction | Doppler frequency shift considered Spatial non-stationary considered, promising for CSI extrapolation in near field. |
| RENEW (M) [180], [181] | Outdoor | 2.4, 5 GHz | Time-frequency-antenna-domain | Static measurement at each site Limit to sub-6 GHz spectrum | Up to 96 antennas beneficial for antenna-domain CSI extrapolation |

TABLE VI
Strengths and limitations of Measured and RT-based Datasets

| Aspects | Measured Datasets | RT-based Datasets |
|---|---|---|
| Realism | High: captures real propagation phenomena, hardware imperfections, noise, and unmodeled effects | Synthetic; realism depends on the accuracy of the 3D scenario model and ray-tracing engine |
| Frequency coverage | Usually limited to one or two bands (mostly sub-6 GHz) | Very wide – Easily covers sub-6 GHz, mmWave, and even sub-THz (up to 100 GHz in WARI-D [178]) |
| Antenna array size | Fixed and generally modest (max 96 antennas in RENEW, 32 in DICHASUS [177]) | Highly scalable – From a few antennas to hundreds/ thousands (DeepMIMO [179], WARI-D [178] up to 384) |
| Mobility | Extremely rare; almost all are static or quasi-static | Almost always static, except DataAI-6G [50] |
| Sampling rate (time domain) | Varies greatly: can be very high (1 kHz) or very low (48 ms in DICHASUS [177]) | No real temporal sampling; users must generate trajectories themselves (except DataAI-6G [50]) |
| Amount of data/ Configurability | Limited by measurement campaign effort; fixed scenarios and positions | Virtually unlimited and fully configurable (geometry, antennas, trajectories, parameters) |
| Subcarrier granularity | Fixed by the hardware | Easily configurable; can generate thousands of subcarriers if desired |
| Cost & Reproducibility | Expensive and time-consuming to collect; hard to reproduce exactly | Low cost after scenario is built; perfectly reproducible and shareable |
| Main strengths | Real-world fidelity, includes practical factors, suitable for validation of models | Broad parameter coverage, massive scale, inclusion of mmWave/THz and near-field effects |
| Main limitations | Narrow frequency/antennas/mobility coverage, limited data volume, static or low-mobility in most cases | Realism limited by scenario accuracy, generally lacks real mobility/Doppler (except rare cases like DataAI-6G [50]), no real hardware impairments |

## VII. Wireless Channel Datasets and Simulators

As machine learning and deep learning are promising to achieve CSI extrapolation, the CSI dataset of high quality plays a vital role in model training. The CSI dataset can be either collected from measurement or generated by simulation. In this section, the open-source datasets and simulators will be reviewed comprehensively.

### A. Open-source channel datasets

The channel measurement of mobile networks is relatively challenging, due to the cost of hardware, and the operators do not open source the channel data due to policy and the value of the data. Fortunately, some institutions and research groups made precious efforts for channel measurement and the resultant channel dataset are summarized in Table V. Current datasets used for evaluation AI/ML in wireless communications can be classified into simulated and measured datasets, which are reviewed as follows:

- Industrial radio dataset [176] is collected through real-world measurements in an industrial factory environment using USRP-based software-defined radios operating at 3.7–3.8 GHz with 80 MHz bandwidth. It provides high-temporal-resolution CSI sampled at 1 kHz, making it one of the few publicly available datasets capable of capturing very fast channel variations. The dataset is particularly suited for time-domain CSI extrapolation and prediction tasks, but mobility is very limited (mostly quasi-static or low-speed scenarios), so it is not representative of high-mobility vehicular or drone channels.

- DICHASUS [177] is a real-world measured massive MIMO dataset gathered in industrial halls at 1.272 GHz center frequency with approximately 50 MHz bandwidth. It features a 32-antenna base station and provides full CSI matrices for 1024 OFDM subcarriers, recorded every 48 ms. Thanks to the large number of subcarriers and antennas, it is excellent for frequency-domain and antenna-domain (spatial) CSI extrapolation research. However, the very low measurement rate (approximately 20.8 Hz) makes it unsuitable for studying fast time-varying channels or Doppler effects.

- Wireless Intelligence[1] is a real-world measured dataset collected on an outdoor university campus at 2.565 GHz with 20 MHz bandwidth using a 4×2 MIMO setup. Similar to the Industrial Radio dataset, it offers a high sampling rate of 1 kHz, enabling fine-grained analysis and extrapolation in the time domain. Measurements were taken at static locations with the receiver fixed at each site, so the dataset captures rich small-scale fading but no significant mobility or Doppler characteristics.

- WARI-D [178] is a large-scale ray-tracing (RT)-generated dataset focused on outdoor scenarios, covering an extremely wide range of frequencies: 2.6 GHz, 6 GHz, 28 GHz, 60 GHz, and 100 GHz. It includes massive antenna arrays (64 elements at lower bands, up to 384 elements at 28 GHz), making it valuable for frequency- and antenna-domain extrapolation across sub-6 GHz, mmWave, and sub-THz bands. Because it is purely static (no user movement), it does not contain temporal or Doppler information.

- DeepMIMO [179] is one of the most widely used RT-based synthetic datasets, supporting both outdoor ("O1" scenario) and indoor ("I1") environments. It provides channel matrices at 3.5 GHz and 60 GHz (with extensions to other bands possible) and is fully configurable: users can arbitrarily set the number of BS/UE antennas, array geometries, and positions. This flexibility makes it ideal for antenna- and frequency-domain extrapolation studies and for generating virtually unlimited training data for machine-learning models. Like most RT-based datasets, channels are static unless the user manually creates trajectories.

- DataAI-6G [50] is a RT-generated dataset designed specifically for dynamic and near-field scenarios, covering outdoor and indoor environments at 3.5 GHz, 6.5 GHz, and 28 GHz. Unlike most other RT datasets, it includes realistic user mobility (UE moving toward the base station in a straight line) with corresponding Doppler shifts and explicitly models spatial non-stationarity, which is crucial for near-field extrapolation and XL-MIMO research. The main drawback is the very small MIMO configuration (1 TX × 4 RX antennas) and the limited movement direction.

- RENEW [180], [181] is a real-world measured dataset collected on the University of California San Diego campus using a large-scale programmable SDR platform operating at 2.4 GHz and 5 GHz. It includes measurements from eight indoor LoS, sixteen indoor NLoS, four outdoor LoS, and twenty-four outdoor NLoS mobile node locations. It supports up to 96 antennas at the base station, providing rich spatial information ideal for antenna-domain CSI extrapolation and massive/XL-MIMO studies. Measurements were performed at static locations, so the dataset contains no significant mobility or high-Doppler components and is limited to sub-6 GHz frequencies.

The measured and RT-based channel datasets are compared comprehensively in Table VI. Specifically, measured datasets offer unparalleled real-world fidelity because they naturally capture complex physical phenomena that remain extremely difficult to model accurately, including diffuse scattering from rough surfaces or vegetation, dynamic blockage and reflections caused by moving people or machinery, near-field spherical wavefronts in very large arrays, antenna mutual coupling, non-ideal radiation patterns, polarization mixing, and a wide range of hardware/RF impairments such as phase noise, IQ imbalance, ADC clipping, and synchronization errors, etc. Despite their realism, measured datasets suffer from severe practical restrictions. Most of the publicly available datasets are confined to sub-6 GHz frequencies and arrays of 128 antennas or fewer in a static or quasi-static manner. Once a measurement is completed, the frequency band, array size, and scenario are fixed; extending the dataset to new bands or hundreds of antennas requires an entirely new, costly measurement effort.

RT-based datasets remove nearly every practical constraint that limits real measurements. A single well-constructed 3D scenario can instantly generate channels from sub-6 GHz all the way to 300 GHz and beyond, support antenna arrays ranging from a few elements to many thousands in arbitrary geometries (planar, cylindrical, spherical, conformal, RIS, etc.), and provide bandwidths from tens of MHz to several GHz. While the realism of RT-based datasets is fundamentally restricted by the accuracy of the underlying 3D scenario and material parameters. Missing objects, incorrect permittivity or conductivity values, or oversimplified vegetation models can introduce systematic errors in certain delay or angular bins, particularly in dense indoor and urban environments. With the notable exception of DataAI-6G, virtually all public RT-based datasets provide only static snapshots or simple scripted trajectories; they lack the rich, random motion of real scatterers (pedestrians, vehicles) that generates continuously evolving Doppler spectra. Finally, RT-based datasets are perfectly clean, i.e., they contain no phase noise, timing offset, nonlinear distortion, or calibration errors.

### B. Channel simulators

Apart from open-soured data set, the academia and industry have made great efforts to simulate channel CSI. The comprehensive review of channel simulators are given in Table VII.

SEU-PML-6GPCS [183], [184] is a 6G pervasive channel model using a unified geometric stochastic channel modeling framework and the integration of 6G full-band full-scene channel characteristics. It supports multi-user, cross-band channel modeling and multiple 3GPP, ITU standardized document. It uses measurement data to calibrate the channel model. Currently, the simulator can implement large number of scenarios, including satellite, UAV [196], terrestrial communication, etc. In addition, for the configuration module, SEU-PML-6GPCS provides menu-driven configuration options for 6G channel parameters, which can be customized to select frequency

TABLE VII

Comparison of channel simulators in terms of key features, whether calibrated with measurements, wether free/open source, supporting spectrum bands and scenarios.

| Simulator | Key features | Calibrated with measurements? | Free/Open source? | Spectrum bands | Scenarios |
|---|---|---|---|---|---|
| SEU-PML-6GPCS [183], [184] | A pervasive GBSM channel model covering 6G full-band full-scene channel characteristics. | Yes, key statistical properties like delay/Doppler/blockage calibrate against measurement channel data. | Yes/No | Sub-6 to optical | 22 standardized 6G use cases with terrestrial/maritime/UAV/LEO |
| BUPTCMG-6G [185]–[188] | GBSM channel model compatible with standard channel models such as ITU-R M.2412, 3GPP TR 38.900/901, TR 36.777, TR 36.873 and TR 37.885. | Yes, undergoing calibration and validation against 3GPP TR 38.901, including key channel parameters such as azimuth spread of arrival, zenith spread of departure, etc. | Yes/No | 0.5 to 330 GHz | Near field, SnS, RIS, NTN, etc. |
| NYUSIM [189] | A measurement-calibrated statistical model supporting MATLAB and python, has been integartted in ns-3 simulator. | Yes, calibrate using extensive real-world measurements for parameters including path loss exponent, delay spread, number of clusters, shadowing, etc. | Yes/Yes | 0.5–150 GHz | UMi/UMa/RMa/InH/InF |
| Sionna RT [189], [190] | GPU-accelerated channel simulator based on TensorFlow; with a unique feature of differentiability, enabling the calculation of gradients for the CIRs, system and environmental-related parameters; further integrated into Omniverse for digital twins. | Yes, via a novel differentiable calibration scheme using channel measurements for material properties scattering behaviors, and antenna patterns. | Yes/Yes | 0.5–300 GHz | 3GPP (UMa, RMa, InH), custom scenarios |
| QuaDRiGa [191], [192] | Quasi-deterministic channel generator, deterministic large-scale paths with stochastic characteristics, compatible with 3GPP TR 36.873, TR 37.885, TR 38.901 TDL, CDL channel models. Could be accelerated by GPU. | Yes, calibrated against 3GPP channel models like 3GPP-3D using extensive channel measurements in Berlin. | Yes/Yes | 0.5–100 GHz | Urban/rural/satellite/V2X/HST |
| WiThRay [193] | High-fidelity 3D RT optimized for RIS and smart environments; accurate EM modeling of reflection/diffraction/scattering on programmable metasurfaces. | Yes, scattering ray calibration for precise EM propagation. | Yes/Yes | sub-6 GHz, mmWave, sub-THz frequency bands | Extremely massive MIMO high mobility |
| NirvaWave [194] | A near-field channel simulator, built on scalar diffraction theory and Fourier principles, providing precise wave propagation response in complex wireless mediums under custom user-defined ransmitted EM signals. | No | No | | Near field with RIS. |
| KUCG [195] | A statistical channel model capable of generating channel impulse responses for millimeter wave and sub-THz bands. | Yes, key statistical parameters calibrated with channel measurements. | Yes/Yes | 60, 95, 105 GHz | Indoor |

bands, scenarios, antenna array sizes and arbitrary motion trajectories of the transceiver and scattering clusters. Then, in the simulation module, SEU-PML-6GPCS can save the full-domain statistical characteristics of the channel, including the time domain, frequency domain, spatial domain, time-delay domain, Doppler domain and angle domain.

BUPTCMCCCMG-IMT2030 [185]–[188] is a link-level channel model simulator for 6G wireless communication. This simulator can generate accurate large scale, small scale parameters and channel coefficients for various scenarios, frequency bands and different antenna arrays. It was officially launched as an implementation of the widely recognized international telecommunication union (ITU)-R M.2412, 3GPP TR 36.873 and 3GPP TR 38.901 standards. It used measurement data to calibrate the channel model. Currently, it can implement standard 5G channel, RIS Assisted channel, IIOT channel and non-terrestrial network channel simulation. Additionally, BUPTCMCCCMG-IMT2030 is implemented using a procedure-oriented framework, including three modules: configuration, simulation, and analysis. In Module I, the configuration module, system parameters like scenario, center frequency, bandwidth and others will be defined. Next in Module II, the simulation module, BUPTCMCCCMG-IMT2030 performs a simulation and generates a series of channel parameters using configurations defined in Module I. Finally in Module III, the analysis module, simulation results are simply analyzed and shown in both numerical and graphical formats.

Developed by NYU WIRELESS, NYUSIM is a widely adopted open-source statistical channel model (available in both MATLAB and Python) that has been integrated into the ns-3 network simulator [189], [190]. It is extensively calibrated using real-world measurement campaigns across multiple bands, accurately capturing path loss exponents, delay and angular spreads, cluster numbers, and shadowing characteristics. Technically, NYUSIM employs a time-cluster and spatial-lobe approach for coupling temporal and spatial domains, generating correlated omnidirectional and directional PDPs based on user trajectories. Version 4.0 introduces multi-polarization support (up to four configurations: V-V, H-H, V-H, H-V) per drop, drop-based simulations for all 3GPP scenarios (UMi, UMa, RMa, InH, InF), and spatial consistency via a four-state Markov model for dynamic human blockage. Frequency support spans 0.5–150 GHz, with scenario coverage including 3GPP UMi, UMa, RMa, InH, and InF environments. Fully open-source. Integration with ns-3 enables end-to-end mmWave/sub-THz network simulations, with computational efficiency improved by restructured APIs and file-based parameter fetching.

NVIDIA's Sionna RT is a GPU-accelerated, differentiable RT engine built on Python [197]. Its unique differentiability enables gradient-based optimization of channel impulse responses with respect to environmental parameters, antenna patterns, and material properties. A novel measurement-driven calibration framework ensures high physical fidelity, using shooting and bouncing rays (SBR) combined with the image method for specular reflections and hashing-based duplicate elimination. It supports specular/diffuse reflections, first-order diffraction, and computes CIRs or radio maps (e.g., path gain, RSS, SINR grids) via heuristic or exhaustive path searches. Integrated with NVIDIA Omniverse, it facilitates digital-twin applications. It supports 0.5–300 GHz and both standard 3GPP scenarios (UMa, RMa, InH) and fully customizable environments. Fully open-source. Version 1.0 overhaul boosts speed (up to 100x via Dr.Jit JIT compiler) and memory efficiency, enabling end-to-end optimization for ISAC, RIS, and ML-based transceivers, with gradients computed for parameters like array geometries and object positions.

QuaDRiGa [191], [192] was developed at Fraunhofer HHI for system-level simulations of mobile radio networks. The supported standardized document are 3GPP TR 36.873, 3GPP TR 37.885 and 3GPP TR 38.901. It was calibrated with measurement data, ensuring its accuracy and reliability in representing real-world propagation characteristics. Additionally, QuaDRiGa supports a variety of application scenarios, including indoor office, indoor industry, UMa, RMa, UMi, V2X and satellite. Besides being a fully-fledged three dimensional geometry-based stochastic channel model, QuaDRiGa contains a collection of features which provides multi-domain features to enable quasi-deterministic multi-link tracking of receiver movements in changing environments.

WiThRay is a high-fidelity 3D RT simulator optimized for RIS-assisted and metasurface-enhanced environments [193]. It provides physically accurate electromagnetic modeling of reflection, diffraction, and scattering on programmable metasurfaces, with calibrated scattering rays for sub-6 GHz, mmWave, and sub-THz bands. Key technical innovations include a bypassing-on-edge (BE) algorithm for efficient path identification, scattering calibration via measurement-based lobe patterns, and geometrical parameter evaluation (e.g., polarization, delay, AoD/AoA) for CIR computation. It generates both continuous CIRs and discrete-time channel data, supporting RIS phase reconfiguration and rough surface scattering. Particularly suited for extremely massive MIMO systems and high-mobility scenarios. Fully open-source. The simulator's EM equation integration ensures ¡1 dB deviation in path gain from theory, enabling evaluations of beamforming, localization, and channel estimation in smart environments with up to $10^6$ rays processed in seconds.

NirvaWave is a specialized near-field channel simulator grounded in scalar diffraction theory and Fourier optics principles [194]. It enables precise computation of wave propagation in complex media under arbitrary user-defined transmitted electromagnetic signals, with native support for near-field RIS scenarios. Technically, it models wavefront evolution (e.g., Airy/Bessel beams) via 2D Fourier-based propagation kernels, incorporating RIS phase shifts and rough scattering for blockage/reflection effects. The core algorithm solves wave equations on user-defined grids, supporting custom TX antenna phase configurations and generating coverage maps for THz/sub-THz bands. Unlike most other tools, it currently lacks explicit measurement-based calibration and is not open-source. Its MIT-licensed implementation (with evaluation editions) achieves orders-of-magnitude runtime efficiency over full Maxwell solvers, facilitating large-scale data generation for model-driven ML techniques in extended near-field regimes.

KUCG is a statistical channel model focused on millimeter-wave and sub-THz indoor environments, supporting 60 GHz, 95 GHz, and 105 GHz bands [195]. Key statistical parameters are calibrated against indoor measurement data. It generates link-level CIRs using cluster-based stochastic modeling, with parameters like path loss, delay spread, and angular spreads derived from propagation measurements. Algorithms include omnidirectional TX antenna assumptions (no AoD generation) and statistical unification for WPAN/WLAN/cellular compatibility. Fully open-source. Version 1.0 evaluation edition emphasizes short-range indoor scenarios, enabling mmWave/sub-THz.

## VIII. Challenges and Research Opportunities

The AI-driven CSI extrapolation presents both significant challenges and exciting opportunities for 6G. This section elaborates the key challenges and potential opportunities in this field.

### A. Challenges of the AI-driven CSI extrapolation for 6G

Although there is extensive research on the AI-driven CSI extrapolation for 6G, limitations of the existing research are significant and are summarized as follows:

*1) Time-domain:* Time-domain CSI extrapolation has traditionally concentrated on short-term prediction horizons [38]. In such cases, spatial stationarity is largely preserved, and Doppler shift caused by UE motion dominates the channel variation. This assumption was adequate for pre-5G systems, in which UE speeds rarely exceeded a few tens of kilometers per hour. However, the emergence of 5G-Advanced and 6G is expected to support scenarios with UE speeds reaching up to 1000 km/h (e.g., high-speed trains). Under these more challenging conditions, long-term extrapolation becomes considerably more difficult: the channel no longer evolves solely because of Doppler effects but also because the UE physically moves into regions with entirely different scattering geometries and propagation environments.

Two fundamental obstacles hinder progress in long-term extrapolation. First, available datasets remain severely limited. Stochastic geometry-based channel models can reproduce Doppler-induced fading caused by motion, yet they struggle to represent large-scale location changes and the strong dependence of channel characteristics on the specific propagation scenario. Conventional TDL and clustered-delay-line (CDL) models defined by 3GPP assume quasi-stationary statistics over short intervals, whereas RT tools excel at position-specific channel reconstruction but cannot easily incorporate continuous UE trajectories. Real-world measurement campaigns can capture both mobility and location-dependent effects, but collecting and annotating such data at scale is prohibitively expensive.

Second, virtually all existing simulation environments treat the propagation surroundings as static. They neglect sudden environmental dynamics, such as moving vehicles, temporary blockages in urban canyons, or swaying foliage, as well as hardware impairments including phase noise and oscillator drift. Consequently, the extrapolation accuracy in the existing research is often overly optimistic. In practical deployments, these unmodeled temporal variations cause rapid degradation of prediction accuracy.

*2) Frequency-domain:* Most prior work on frequency-domain CSI extrapolation has targeted FDD systems operating within the same broad spectrum band (typically sub-6 GHz). In these configurations, the UL and DL carrier frequencies are close enough that partial channel reciprocity is preserved: the primary difference between UL and DL lies in the complex-valued. This limited discrepancy greatly simplifies the CSI extrapolation for FDD systems.

In contrast, CSI extrapolation across widely separated frequency bands, such as between sub-6 GHz and mmWave, or across different mmWave segments, poses a far greater challenge. Path loss exponents, multipath richness and scattering mechanisms, etc, all vary considerably with frequency. As a result, single-band extrapolation techniques, cannot be straightforwardly applied. Establishing accurate bidirectional mappings is particularly difficult because the relationship between channel responses at distant frequencies is inherently nonlinear and band-specific. Errors introduced during extrapolation from a source band therefore tend to accumulate rapidly when the process is reversed or extended to additional bands, leading to significant performance degradation.

*3) Antenna-domain:* Most studies on antenna-domain CSI extrapolation have concentrated on conventional MIMO systems operating in the far-field regime. In such setups, the wavefronts are essentially planar across the entire antenna array, so every transmit-receive antenna pair experiences nearly identical propagation paths determined solely by angle. Emerging paradigms such as XL-MIMO, RIS, and FAS, operate partly or wholly in the near-field region. Here, spherical wavefronts become significant, and the visibility of individual scattering paths varies across the array: certain multipath components may illuminate only a subset of antenna elements. This spatial non-stationarity fundamentally complicates antenna-domain extrapolation.

Two primary obstacles limit progress in this area. First, acquiring realistic near-field channel datasets remains extraordinarily difficult. Measurement campaigns that capture full spherical-wave effects over large-aperture arrays are costly, time-consuming, and require precise positioning of both trans-mitter and receiver. To date, no widely available channel simulator can accurately reproduce these near-field phenomena with sufficient fidelity for training or evaluation. Second, the inherent non-stationarity of near-field channels due to the change of path visibility across the Tx-Rx pairs is hard to model and predict. Existing extrapolation methods, which typically assume smooth spatial correlation, struggle to track these discontinuities, resulting in rapid performance degradation.

*4) Multi-domain:* Multi-domain CSI extrapolation, which jointly operates across the time, frequency, and antenna-domain, magnifies the limitations discussed earlier. Errors and simplifying assumptions that are tolerable within a single domain no longer remain isolated; instead, they interact and amplify across domains, producing far larger deviations from reality.

Even advanced channel simulators such as QuaDRiGa or Sionna can only partially address this challenge. Although they model certain cross-domain correlations under controlled conditions, they still rely on idealized propagation assumptions that fail to capture the full complexity of real-world scenarios. Examples include the tightly coupled effects of high-speed mobility and large frequency separation in EL-MIMO systems (e.g., drone-mounted or low-Earth-orbit satellite links), or the simultaneous impact of weather-dependent scattering (such as rain-induced attenuation and depolarization) on temporal, spectral, and spatial channel characteristics. As a result, extrapolation models trained or evaluated solely on simulated data often achieve deceptively good performance in individual domains yet suffer from severe joint prediction errors when deployed in practical environments.

This gap underscores the urgent need for hybrid datasets that strategically combine high-fidelity RT or stochastic simulations with targeted measurement campaigns. Only such composite datasets can provide the diverse, physically consistent multi-domain samples required to train robust and generalizable extrapolation frameworks.

### B. Future research

Based on the analysis of the limitations of the existing research in CSI extrapolation, we propose the following future research.

*1) Effective and reliable dataset construction:* High-quality dataset with massive volume is critical for AI-driven CSI extrapolation, the following future research can be considered to overcome the scarcity of high-quality dataset.

- **Exploiting existing datasets**:
  Effectively leveraging existing datasets is a critical approach to addressing the scarcity of high-quality data for CSI extrapolation tasks. However, available datasets in wireless channel research are predominantly simulation-based, with real-world measurement data being scarce. Moreover, datasets from different sources often vary in format, scenarios, and parameter configurations, limiting their direct applicability to CSI extrapolation. *a)* Dataset alignment techniques can standardize these datasets by unifying key parameters such as frequency, bandwidth, and antenna configurations, thereby enhancing their reusability [198]. Specifically, a potential approach is developing an open-source data processing framework that integrates data cleaning, format conversion, and parameter alignment functionalities [199], [200]. This framework would enable the unification of heterogeneous channel datasets (e.g., RT-generated or measured data) into a standardized format. It is expected to automatically correct frequency offsets and antenna configuration discrepancies, coupled with statistical validation modules, would ensure data consistency. *b)* Additionally, the generation of synthetic datasets is also a promising direction. By employing generative models such as GANs [182], [201] and diffusion models [202], synthetic channel data with diversity and representativeness can be created to augment limited real-world datasets. Specifically, conditional GANs (CGANs) are promising to generate scenario-specific channel data, conditioned on parameters such as frequency, distance, and mobility speed, as well as statistical characteristics of existing datasets. These synthetic datasets must undergo rigorous statistical validation (e.g., path loss, Doppler spread) to ensure their distribution characteristics align with actual channel environments, thus supporting the training and evaluation of CSI extrapolation models.

- **Developing effective channel measurement**:
The development of cost-effective and efficient channel measurement platforms is a fundamental solution to the shortage of high-quality datasets. Traditional channel measurement equipment, such as vector network analyzers or dedicated channel sounders, is expensive and complex to operate, limiting large-scale data collection. *a)* A promising solution is developing low-cost measurement platforms based on Software-Defined Radio (SDR), such as USRP or HackRF, and using Commercial Off-The-Shelf (COTS) hardware and open-source software, such as MATLAB or GNU Radio for signal processing and channel parameter extraction [203], [204]. Such platforms should be further extended to support measurements across multiple frequency bands (e.g., Sub-6 GHz, millimeter-wave, or THz) and enable the extraction of multi-antenna channel parameters. Also, a user-friendly interface should be developed to reduce operational complexity. *b)* Low-cost UAVs have been employed for flexible channel measurements [205], [206], and Machine learning-driven measurement strategies can further enhance data collection efficiency and quality by predicting the spatial distribution of channel parameters, guiding optimal measurement point selection to minimize redundancy and maximize scenario coverage. Specifically, reinforcement learning algorithms could be used to develop adaptive path-planning tools for measurements. By leveraging environmental information and historical data, these tools can predict channel variations and optimize drone trajectories to maximize scenario coverage and parameter diversity.

- **Advancing Channel Simulator**:
Developing advanced channel simulators is a vital strategy for improving data availability in CSI extrapolation research. Current simulators, such as geometry-based (e.g., RT) or statistical models, often fail to capture the non-stationary characteristics or higher-order statistical properties of complex wireless channels. Future research can focus on hybrid channel simulators that integrate geometry-based models (e.g., RT) with data-driven approaches (e.g., NNs) to more accurately replicate the dynamic behavior of wireless channels. *a)* For instance, a hybrid simulator has been proposed in [50], which could model the near-field effect and Doppler shift based on the simulation results of RT. Further research should focus on integrating more factors in 6G, such as the RIS, fluid antenna systems, mmWave, THz band, UAV [207], etc. *b)* Moreover, the real-time capability and scalability of channel simulators are critical areas of development. A potential solution is building a distributed channel simulation system on cloud platforms (e.g., AWS or Azure) using GPU acceleration to enable real-time generation of large-scale MIMO or THz channel data. The platform should provide API interfaces to facilitate integration into the development pipeline of CSI extrapolation algorithms.

### 2) Comprehensive Performance Evaluation:

- **Generalization Performance**:
The generalization capability of CSI extrapolation models represents a critical metric for assessing their practical deployment performance, particularly in dynamic and unseen scenarios. In addition, significant differences in measurement setups, including calibration procedures, antenna polarization, transmit power, acquisition duration, and local scattering environments, often lead to severe distribution shifts when datasets from different campaigns are naively combined. Such shifts can result in over- or under-estimation of true generalization performance. How to fairly evaluate the generalization performance remains an open problem, potential future research directions are summarized as follows: *a)* Scenario heterogeneity index: a statistical distance (e.g., KL divergence) computed over key scenario parameters (center frequency, bandwidth, array aperture, environment type, mobility, etc.) to quantify the discrepancy between source and target domains. *b)* Blockage/anomaly prediction score: A dedicated metric for dynamic blockage or anomalous events (e.g., sudden occlusion by vehicles or pedestrians), combining F1-score, prediction lead time, and false alarm rate. *c)* Transferability ratio: The ratio of zero-shot performance to few-shot fine-tuned performance on the target domain; a value closer to 1 indicates stronger intrinsic generalization.

- **Dataset volume required for domain-specific CSI extrapolation**:
One of the fundamental questions for AI-driven CSI extrapolation is the volume of dataset required in each domain to achieve desirable performance. Intuitively, the volume of dataset required depends on the AI model, the complexity and dynamics of scenarios, for example, self-supervised learning generally requires much less data than supervised learning [157], and more data is needed for NLoS than LoS scenarios [51]. However, there is limited research on this topic in wireless communications, especially the quantitative results regarding the dataset volume for each domain. *a*) [22] observe that longer historical CSI input enhances the time-domain CSI extrapolation using Transformer-based models due to their capability of capturing long-term temporal correlation, while the length of historical CSI input sequence has much limited influence for LSTM/GRU-based models. [57] observed the performance of CSI extrapolation in antenna-domain is improved by increasing the number of input CSIs and saturate at certain number of inputs. However, all the above research is qualitative, and lacks quantitative analysis. In addition, as the simulation settings for existing research on CSI extrapolation are generally distinct, it is challenging to give solid answers by literature review. To this end, we propose to build a public dataset, like the ImageNet for computer vision, which could be used as a benchmark to quantitatively evaluate the balance between historical information and prediction horizon. 3GPP has made pioneering effects to build such dataset, but requires further dedication from whole community. *b*) Inspired by the filed of AI, one promising approach to find out the minimum dataset volume is active learning, which selects the most helpful dataset for model training under the metrics of uncertainty and diversity [208]. Uncertainty quantifies how challenging the data samples are for downstream tasks, while diversity measures how different the data samples are compared to the dataset has been used for training. However, how to compute similar metrics for channel data is much challenging due to its high dimensionality, including scenario features, carrier frequency, velocity, antenna spacing, etc.

- **Computation-efficient model design:**
The computational complexity of AI-based CSI extrapolation models is a critical factor for their deployment in resource-constrained environments, such as edge devices or real-time 6G applications, yet existing evaluation metrics often overlook this aspect [209]. High-complexity models, such as Transformers or deep convolutional neural networks, may achieve superior extrapolation accuracy but incur significant computational costs, making them impractical for latency-sensitive scenarios like V2X communication. *a*) Sparse Attention Transformers offer significant advantages for low-complexity CSI extrapolation by reducing the quadratic computational complexity of traditional self-attention mechanisms to near-linear or logarithmic levels, making them suitable for processing high-dimensional CSI in real-time applications like 6G networks [209], [210]. Techniques such as Routing Transformer or Performer leverage sparse attention patterns (e.g., local or linearized attention) to focus computational resources on relevant CSI features, such as spatial correlations or Doppler shifts, thereby enhancing extrapolation accuracy while minimizing latency and memory usage. This efficiency is particularly beneficial for massive MIMO or millimeter-wave scenarios, where CSI sequences are long and complex, enabling robust predictions across diverse environments (e.g., urban, vehicular). *b*) Lightweight CNNs, such as MobileNetV2 or EfficientNet, provide a compelling solution for low-complexity CSI extrapolation by leveraging techniques like depthwise separable convolutions, model pruning, and quantization to minimize computational cost and memory footprint [211], [212]. These models are well-suited for resource-constrained devices, such as IoT nodes or mobile terminals, where they can efficiently process CSI to predict channel responses in dynamic environments, such as vehicular or UAV networks [213]–[215]. Their ability to capture spatial-frequency correlations in CSI with fewer parameters than traditional CNNs ensures high extrapolation accuracy while meeting the low-latency requirements of 6G applications. Knowledge distillation from larger models further enhances their performance, making them viable for edge computing. However, research challenges include the limited expressive power of lightweight CNNs compared to deeper architectures, which may struggle to model complex channel dynamics (e.g., non-stationary scenarios). Designing lightweight CNNs that generalize across diverse channel scenarios requires careful architecture optimization and data augmentation strategies.

*3) Advanced model design:*

- **Model-Driven CSI extrapolation:**
The robustness of AI algorithms in wireless communication systems is often challenged by the noise, dynamic and complex nature of channel environments. Model-driven approaches, which combine domain-specific physical models with data-driven techniques, have shown significant promise in enhancing the resilience and generalization of CSI extrapolation models [216], [217]. These methods leverage prior knowledge of wireless channel characteristics, such as path loss, multipath fading, and Doppler effects, to guide the learning process, thereby reducing reliance on extensive labeled datasets. Specifically, physics-informed neural networks could embed physical channel models (e.g., Saleh-Valenzuela model) into the loss function of DNNs and can optimize channel parameters like path loss and delay spread while learning from data, ensuring robust extrapolation across frequency bands and environments. Future research should focus on seamlessly integrating model-driven and data-driven paradigms to create hybrid frameworks [218]. For example, iterative parameter estimation algorithms traditionally used for channel modeling can be reformulated as DNNs, where physical channel parameters (e.g., AoA, delay spread) are embedded as learnable features. This approach expands the model's parameter space while allowing fine-tuning through data-driven optimization, significantly improving extrapolation accuracy and robustness at different noise levels.
- **Novel Architecture Design:** The transformative potential of Transformer-based architectures in AI-driven CSI extrapolation is increasingly evident, positioning them as a cornerstone for future model development [219], [220]. However, their computational complexity and resource demands pose challenges for practical deployment in wireless systems. To address these, sparse attention mechanisms and mixture-of-experts (MoE) architectures offer promising solutions to enhance extrapolation accuracy, generalization, and computational efficiency. Sparse attention techniques, such as those employed in the Routing Transformer [209] or hierarchical attention models [221], reduce the quadratic complexity of self-attention by focusing on relevant feature interactions, making them suitable for processing long-sequence or high-dimensional channel data. Similarly, MoE architectures, as demonstrated in Switch Transformers [222] and GShard [223], introduce sparsity in feed-forward layers by selectively activating a subset of experts, enabling scalable and efficient large-scale models.

- **Multi-Modal Assisted CSI extrapolation:** Incorporating multi-modal data has emerged as one of the most effective strategies for overcoming the inherent limitations of CSI-only extrapolation. Wireless channel characteristics are influenced by diverse factors, including environmental geometry, user mobility, and sudden blockages, etc., which cannot be fully captured by a single data modality (e.g., CSI). Multi-modal learning addresses this gap by fusing complementary sensors and data streams, including camera images [224], [225], mmWave radar point clouds [226], WiGig/Wi-Fi sensing side-information [227], [228], or even publicly available maps and weather feeds. When properly integrated, these sources provide rich information about the propagation environment [229] that is otherwise implicit or entirely absent in the received signal, dramatically improving spatial-temporal prediction accuracy in challenging settings such as dense urban canyons, indoor factories, or high-mobility outdoor scenarios. A key challenge in multi-modal CSI extrapolation is designing fusion mechanisms that effectively combine heterogeneous data while mitigating issues such as modality misalignment or data scarcity. Advanced techniques, such as cross-modal attention or GNNs, can be employed to model relationships between modalities, enabling the model to learn robust and generalizable channel representations.

*4) Integrating with Emerging Techniques:* CSI extrapolation is able to acquire CSI with low overhead, thereby promising to enhance the performance of emerging technologies in 6G. In this section, we will elaborate the opportunities and challenges of integrating CSI extrapolation with emerging technologies, and discuss potential future research.

- **Emerging MIMO systems:** Advanced antenna technologies, XL-MIMO, RIS and FAS, are fundamentally reshaping wireless systems and placing unprecedented demands on CSI extrapolation methods [230]–[232]. In XL-MIMO, spatial non-stationarity implies that different portions of the aperture observe distinct sets of scattering paths, with considerable variations in path number, gain, phase, and delay across the array [49]. RIS and FAS further introduce controllable yet rapidly time-varying channels: RIS through programmable reflection coefficients and FAS through dynamic port relocation. These characteristics dramatically increase the dimensionality and temporal volatility of the channel, making real-time extrapolation far more demanding, especially on power- or compute-constrained user devices [232], [233]. Designing AI models capable of operating accurately under these conditions while respecting strict latency and energy constraints remains an open challenge. The publicly available, standardized datasets that adequately capture the physics of these emerging technologies is also critical to accelerate the research on this field.

- **Emerging Multi-Access Technologies:** Emerging multiple-access techniques, particularly non-orthogonal multiple access (NOMA) [234] and rate-splitting multiple access (RSMA), have become cornerstone enablers for the spectral efficiency and massive connectivity targets of 6G networks [235], [236]. Unlike orthogonal schemes, both approaches deliberately introduce controlled interference among users, which places significantly higher demands on the accuracy and timeliness of CSI extrapolation. *a)* In NOMA, several users share the same time-frequency resources through power-domain or code-domain multiplexing. Successful decoding at the receiver, and optimal power allocation at the transmitter depend heavily on precise knowledge of instantaneous inter-user channel conditions. When perfect CSI is unavailable, extrapolation errors directly transfer into residual interference, degraded successive interference cancellation, and reduced throughput. This problem is especially pronounced in highly dynamic environments such as vehicular networks, where AI-based time-domain channel extrapolation must anticipate rapid channel variations while maintaining low latency. *b)* RSMA pushes complexity further by splitting each user's message into a common stream (decoded by multiple receivers) and a private stream (treated as noise by others). The resulting interference landscape is a hybrid of broadcast and unicast characteristics, requiring the extrapolation framework to simultaneously track channel responses for both stream types and their mutual dependencies. AI-driven extrapolation can address this by jointly modeling the channel responses for common and private streams, leveraging multi-task learning to capture their inter-dependencies.

- **Satellite communications:** Satellite communications introduce distinctive challenges for CSI extrapolation that have sparked growing research interest in long-term, robust prediction techniques [14], [92], [96], [97], [103], [107], [237]. Two factors dominate the difficulty: extreme user mobility, especially in low-Earth-orbit (LEO) constellations and high-speed aerial or maritime platforms, and round-trip delays that range from tens to hundreds of milliseconds. These conditions demand extrapolation horizons far beyond those typical of terrestrial systems, often requiring accurate forecasts hundreds of milliseconds into the future despite rapidly evolving Doppler, multipath, and blockage dynamics [238]. Processing long prediction sequences on board resource-constrained terminals imposes strict limits on model complexity and power consumption. Publicly available, high-fidelity satellite channel datasets remain extremely scarce, forcing researchers to rely on simplified geometric-stochastic models or domain-adapted transfer learning from terrestrial datasets. Finally, the highly non-stationary nature of the channel—driven by orbital motion, varying elevation angles, and atmospheric turbulence—makes conventional metrics inadequate. A promising solution is exploiting multimodal inputs that are readily available in satellite systems, such as precise satellite ephemeris, user trajectory forecasts, and real-time ionospheric or tropospheric scintillation indices to improve CSI extrapolation [239].

## IX. Conclusions

CSI extrapolation techniques play a crucial role in modern wireless communication systems, especially in the upcoming 6G era. With the rapid growth of communication demand and the increasing requirements of high SE and low latency, traditional channel estimation methods are no longer able to cope with the challenges posed by high-dimensional data processing and emerging technologies. CSI extrapolation techniques infer the complete CSI by using partial CSI, thus effectively reducing the frequent feedback and transmission overheads and improving the efficiency of system resource utilization.

However, there is still a lack of comprehensive survey on the existing research in CSI extrapolation. To this end, we review the research on CSI extrapolation in-depth in this paper. Specifically, we first introduce the fundamentals of CSI extrapolation in time, frequency, antenna and multi-domain, and representative AI models. We then review the existing research on CSI extrapolation in time, frequency, antenna and multi-domain comprehensively covering the principles of major techniques along with their strengths and weaknesses. As the key resource for the era of AI, the open-source CSI datasets and channel simulator are summarized. To resolve the challenges of CSI extrapolation in time, frequency, antenna and multi-domain, we also provide promising future research directions covering the dataset construction, performance evaluation, model design and integrating with emerging technologies in 6G.

To conclude, CSI extrapolation technology has become an indispensable part of future wireless communication systems, and its research and development will not only help to improve the performance of communication systems, but also lay a solid foundation for realizing the vision of 6G networks.